\documentclass{elsarticle}

\usepackage{hyperref}
\usepackage[margin=0.9in]{geometry}

\journal{Neurocomputing}

\usepackage{framed,multirow}
\usepackage{amsmath,amssymb,amsfonts}
\usepackage{algorithmic}
\usepackage{graphicx}
\usepackage{textcomp}
\usepackage{amssymb}
\usepackage{latexsym}
\usepackage[labelformat=simple]{subfig}
\usepackage{url}
\usepackage{color}

\usepackage[figuresright]{rotating}

\begin{document}

\begin{frontmatter}




\title{CyTran: A Cycle-Consistent Transformer with
Multi-Level Consistency for Non-Contrast to Contrast CT Translation}


\author[1,2,3]{Nicolae-C\u{a}t\u{a}lin {Ristea}}
\author[4,5]{Andreea-Iuliana {Miron}}
\author[4,5]{Olivian {Savencu}}
\author[1]{Mariana-Iuliana {Georgescu}}
\author[4,5]{Nicolae {Verga}}
\author[3,6]{Fahad Shahbaz {Khan}}
\author[1,7]{Radu Tudor {Ionescu} \corref{cor1}}

\address[1]{Department of Computer Science, University of Bucharest, Romania}

\address[2]{Department of Telecommunications, University Politehnica of Bucharest, Romania}
\address[3]{Mohamed bin Zayed University of Artificial Intelligence, UAE}
\address[4]{Col\c{t}ea Hospital, Romania}
\address[5]{Carol Davila University of Medicine and Pharmacy, Romania}
\address[6]{CVL, Linköping University, Sweden}
\address[7]{Romanian Young Academy, University of Bucharest, Romania}

\cortext[cor1]{Corresponding author: 
  E-mail: raducu.ionescu@gmail.com}

\begin{abstract}
We propose a novel approach to translate unpaired contrast computed tomography (CT) scans to non-contrast CT scans and the other way around. Solving this task has two important applications: (i) to automatically generate contrast CT scans for patients for whom injecting contrast substance is not an option, and (ii) to enhance the alignment between contrast and non-contrast CT by reducing the differences induced by the contrast substance before registration. 

Our approach is based on \emph{cy}cle-consistent generative adversarial convolutional \emph{tran}sformers, for short, \emph{CyTran}. Our neural model can be trained on unpaired images, due to the integration of a multi-level cycle-consistency loss. Aside from the standard cycle-consistency loss applied at the image level, we propose to apply additional cycle-consistency losses between intermediate feature representations, which enforces the model to be cycle-consistent at multiple representations levels, leading to superior results. To deal with high-resolution images, we design a hybrid architecture based on convolutional and multi-head attention layers. In addition, we introduce a novel data set, \emph{Coltea-Lung-CT-100W}, containing 100 3D triphasic lung CT scans (with a total of 37,290 images) collected from 100 female patients (there is one examination per patient). Each scan contains three phases (non-contrast, early portal venous, and late arterial), allowing us to perform experiments to compare our novel approach with state-of-the-art methods for image style transfer. 

Our empirical results show that CyTran outperforms all competing methods. Moreover, we show that CyTran can be employed as a preliminary step to improve a state-of-the-art medical image alignment method. We release our novel model and data set as open source at: \url{https://github.com/ristea/cycle-transformer}.

Our qualitative and subjective human evaluations reveal that CyTran is the only approach that does not introduce visual artifacts during the translation process. We believe this is a key advantage in our application domain, where medical images need to precisely represent the scanned body parts.
\end{abstract}

\begin{keyword}
Transformers\sep
generative adversarial transformers\sep 
deep learning\sep
cycle-consistency\sep 
image translation\sep
image registration\sep
computed tomography\sep
triphasic lung CT.
\end{keyword}

\end{frontmatter}


\section{Introduction}
\label{sec:introduction}

Patients undergoing computed tomography (CT) screening may not be able to get intravenous injections with contrast agents due to allergies~\cite{Namasivayam-ER-2006} or other medical conditions, e.g.~muscular dystrophy, causing low blood creatinine levels. However, the contrast substance plays a very important role in helping medical experts to detect and delimit certain lesions, e.g.~malignant tumors \cite{Yan-OT-2017}. For instance, radiotherapy heavily relies on the accurate segmentation of tumors.

Certainly, the majority of patients are healthy enough to be injected with contrast agents, enabling medical experts to obtain triple-phase CT scans which provide a significantly more clear picture of the malignant lesions. The triple-phase CT includes a native phase (before the contrast is injected), a portal venous phase (right after the contrast is injected), and a late arterial phase (when the contrast passes into the arteries). As the three phases are in successive temporal order, the corresponding CT scans are taken in different moments in time, inherently leading to a misalignment between scans caused by slight movements of the patient, e.g.~due to breathing. In this scenario, some image registration method can be employed to align the CT scans. However, due to Hounsfield unit (HU) differences between contrast and non-contrast CT in certain types of tissue, the image alignment task becomes problematic, especially for the regions of interest, where tumors are located.

A method capable of translating between contrast and non-contrast CT scans, in both directions, is likely to solve the two problems presented above. Indeed, when patients are unable to undergo contrast CT screening, the medical experts could simply employ the automated translation technique to generate contrast CT scans corresponding to the venous or arterial phases. When CT scans need to be aligned, the same translation technique can be applied prior to the alignment step to remove the effect of the contrast substance, thus allowing for a better alignment. Once the displacement field is generated, it can be applied on the unmodified contrast CT scan to obtain the final alignment result.

In this work, we present a novel deep learning model to translate between contrast and non-contrast CT scans. Our method relies on the success of generative adversarial networks (GANs) \cite{Goodfellow-NIPS-2014} in image-to-image translation \cite{Zhu-ICCV-2017}, a class of vision tasks where the goal is to learn the mapping between an input image and an output image using a training set of paired or unpaired images from two different domains. Although GANs have been previously employed for medical image-to-image translation \cite{Armanious-EUSIPCO-2019, Chandrashekar-EHJ-2020, Kearney-RAI-2020, Pengjiang-JGP-2020}, we underline that the task of translating between contrast and non-contrast CT scans is very challenging due to the requirement of recognizing specific tissue types, anatomical structures, and even tumors, which exhibit significant HU changes between contrast phases. Failing to recognize such anatomical structures raises the possibility of introducing unrealistic information. This would evidently lead to unreliable synthetic images, which cannot be used for diagnosis or treatment purposes. Convolutional neural networks fail to recognize the global structure of objects in natural images \cite{Sabour-NIPS-2017}, and, for the same reason, can fail to recognize anatomical structures in CT scans. To this end, we propose a \textbf{cy}cle-consistent generative adversarial \textbf{tran}sformer, called \emph{CyTran}, which has a higher capacity of recognizing global structure due to the incorporated transformer block. Since pure vision transformers cannot process high-resolution images due to the high number of learnable parameters involved, we design a hybrid architecture comprising both convolutions and multi-head attention, such that it can generate full-resolution CT scans. To avoid mode collapse, we apply a cycle-consistency loss between each input image and the image translated back to its original domain, following Zhu et al.~\cite{Zhu-ICCV-2017}. Along with the standard cycle-consistency loss applied at the image level, we introduce two additional cycle-consistency losses between intermediate feature representations, enforcing the model to be cycle-consistent at multiple representations levels.

To demonstrate the applicability of CyTran on the important cases exemplified above, we introduce the \emph{Coltea-Lung-CT-100W} data set formed of 100 3D anonymized triphasic lung CT scans of female patients. For each patient, there are three CT scans corresponding to the native, venous and arterial phases, respectively. We conduct a set of experiments to compare CyTran with other state-of-the-art style transfer methods, namely CycleGAN \cite{Zhu-ICCV-2017}, pix2pix \cite{Isola-CVPR-2017}, U-GAT-IT \cite{Kim-ICLR-2019}, CWT-GAN \cite{Lai-ICCV-2021} and AttentionGAN \cite{Tang-TNNLS-2021}. Both automatic and human evaluations indicate that our approach is consistently better than the competing methods. We perform another set of experiments to evaluate the applicability of style transfer methods in contrast to non-contrast CT alignment with a state-of-the-art 3D image registration method \cite{Chen-AIM-2020}. While the empirical results indicate that all style transfer methods are helpful, the highest performance improvements are brought by CyTran, confirming the superiority of our approach.

To the best of our knowledge, our contribution is fourfold:
\begin{itemize}
    \item We design a cycle-consistent generative adversarial transformer in medical imaging, demonstrating state-of-the-art results in style transfer between contrast and non-contrast CT scans.
    \item We propose to introduce cycle-consistency losses at different feature representation levels, improving the quality of the generated images.
    \item We publicly release a novel data set containing triphasic lung CT scans.
    \item We employ style transfer methods to enhance the alignment between contrast and non-contrast CT scans.
\end{itemize}

\section{Related Work}

\subsection{Transformers}

Architectures based on self-attention, in particular transformers \cite{Vaswani-NIPS-2017}, have become the model of choice in natural language processing (NLP). Thanks to the computational power and scalability of transformers, it has become possible to train models of unprecedented size \cite{Brown-NIPS-2020}. With the ever growing size of models and data sets, the performance improvements are constantly increasing. Considering the success of transformers in NLP \cite{Brown-NIPS-2020}, architectures based on multi-head self-attention have been adopted in the computer vision community, attaining superior results in various tasks \cite{Dosovitskiy-ICLR-2020, Khan-arXiv-2021, Wu-arXiv-2021, Zhang-arXiv-2021}.
Wu et al.~\cite{Wu-arXiv-2021} proposed a convolutional transformer that improves the Vision Transformer (ViT) \cite{Dosovitskiy-ICLR-2020} in terms of performance and efficiency by introducing convolutions into the model, in an attempt to take advantage of both designs. We further replace the multi-layer perceptron from the transformer block proposed by Wu et al.~\cite{Wu-arXiv-2021} with pointwise convolutional layers, allowing us to incorporate the resulting block into an efficient generative model that can produce high-resolution images. Zhang et al.~\cite{Zhang-arXiv-2021} proposed a cycle-consistent attention mechanism for semantic segmentation. In contrast, we introduce a novel generative model based on an efficient convolutional transformer backbone, where the cycle consistency is imposed at multiple representation levels, not only at the latent feature level.

In medical imaging, the popularity of transformer architectures is rising \cite{Chen-arXiv-2021, Gao-arXiv-2021, Hatamizadeh-arXiv-2021, Korkmaz-arXiv-2021, Luthra-arXiv-2021}, most likely because such models bring state-of-the-art results. 
For instance, Gao et al.~\cite{Gao-arXiv-2021} proposed an efficient self-attention mechanism which reduces computational complexity for cardiac magnetic resonance imaging (MRI) segmentation. Korkmaz et al.~\cite{Korkmaz-arXiv-2021} presented a zero-shot learning method employing a cross-attention transformer block to reconstruct MRI images. Luthra et al.~\cite{Luthra-arXiv-2021} proposed an encoder-decoder network, which uses transformer blocks for CT image denoising. Different from the medical imaging methods based on transformers, we introduce a novel convolutional transformer architecture for style transfer between contrast and non-contrast CT images, which can be trained on unpaired data through the use of multiple cycle-consistency terms in the loss. To our knowledge, this is the first work to propose cycle-consistent transformers in medical imaging.

\subsection{Image Translation}

Since the introduction of GANs \cite{Goodfellow-NIPS-2014} in 2014, a large body of research has focused on theoretical and architectural changes \cite{Gulrajani-NIPS-2017, Isola-CVPR-2017, Salimans-ANIPS-2016, Soviany-WACV-2020, Zhu-ICCV-2017}, giving rise to a wide adoption of GANs across various generative tasks, including image translation \cite{Isola-CVPR-2017,Kim-ICLR-2019,Lai-ICCV-2021,Tang-TNNLS-2021,Zhu-ICCV-2017}. 
In 2016, the pix2pix framework \cite{Isola-CVPR-2017} became one of the first GAN models to address the task of image-to-image translation from a source domain image, e.g.~a springtime landscape, to a corresponding target domain image, e.g.~a winter landscape, provided that paired images from the two different domains are available for training. To overpass the lack of paired data sets for style transfer, researchers have developed methods for unpaired image-to-image translation \cite{Zhu-ICCV-2017}. Zhu et al.~\cite{Zhu-ICCV-2017} solved the problem by using two generators, one that translates a source image to the target domain, and the other to translate the translated image back to the source domain. The two generators are optimized such that the image passing through the two generators is close to the original input image, ensuring the cycle-consistency of the framework. More recently, Lai et al.~\cite{Lai-ICCV-2021} introduced a cross-model weight transfer mechanism to transfer a certain proportion of the weights from the discriminator to the generator, after each training iteration. Tang et al.~\cite{Tang-TNNLS-2021} observed that state-of-the-art GANs for unpaired image translation still generate visual artifacts, proposing to alleviate this problem by training attention-guided generators to produce attention masks that are fused with the generated image, thus increasing image quality. They introduced attention in the discriminator as well, ensuring that it focuses on attended regions.

GANs have also been adopted in medical imagining, often being employed for medical image translation \cite{Choi-JNM-2018,Emami-MP-2018,Kearney-RAI-2020, Modanwal-MI-2020,Seo-WCACV-2021,Wolterink-TMI-2017}. For example, Seo et al.~\cite{Seo-WCACV-2021} proposed a two-stage algorithm to address style transfer between contrast and non-contrast CT images. The first stage removes the poor alignment effects, while the second stage relies on a GAN architecture to enhance the contrast of CT images. Other approaches used the pix2pix framework in applications where paired images are available, such as positron emission tomography (PET) to MRI translation \cite{Choi-JNM-2018}, organ segmentation \cite{Huo-MI-2018}, MRI to CT translation \cite{Emami-MP-2018}, and low-dose CT denoising \cite{Wolterink-TMI-2017}. More recently, researchers started using CycleGANs \cite{Zhu-ICCV-2017} for various medical imaging tasks \cite{Kearney-RAI-2020, Modanwal-MI-2020}. For instance, Kearney et al.~\cite{Kearney-RAI-2020} employed a CycleGAN \cite{Zhu-ICCV-2017} to translate between MRI and CT data. Modanwal et al.~\cite{Modanwal-MI-2020} proposed an algorithm that modifies CycleGAN by introducing two discriminators to translate between different MRI images. Closer to our task, Chandrashekar et al.~\cite{Chandrashekar-EHJ-2020} proposed an algorithm which relies on CycleGAN to enhance the contrast of CT images. To the best of our knowledge, none of the related methods are based on transformer architectures. We provide empirical evidence showing that cycle-consistent transformers outperform architectures based on pix2pix or CycleGAN when it comes to contrast to non-contrast CT translation and back. Furthermore, to the best of our knowledge, the idea of optimizing features at multiple levels to become cycle-consistent has not been explored so far in medical imaging.

\subsection{Image Registration}

Medical image registration is a fundamental problem that improves visual inspection, diagnosis, and treatment planning. It refers to the task of establishing spatial correspondences between points in a pair of fixed and moving images through a spatially varying deformation model. The state-of-the-art methods for medical image alignment are based on deep neural networks \cite{Balakrishnan-CVPR-2018,Burduja-ICIP-2021,Chen-VITVN-2021,Krebs-MICCAI-2017, Rohe-MICCAI-2017,Zhao-JBHI-2019, Zhao-ICCV-2019}. A recent registration approach proposed in \cite{Zhao-ICCV-2019} is based on a recursive cascade algorithm which assumes that, at each step, the neural model learns to perform a progressive deformation of the current warped image. Nevertheless, the trend of applying transformers has recently been adopted in medical image registration as well. However, Chen et al.~\cite{Chen-VITVN-2021} claimed that architectures based solely on transformers emphasize the low-resolution features because of the consecutive downsampling operations, resulting in a lack of detailed localization information that affects image registration performance. To alleviate this problem, the authors combined transformers with convolutional layers into an architecture called ViT-V-Net, which improves the recovery of localization information. 

Different from other medical image registration approaches, we employ CyTran as a data augmentation method to improve alignment results. The augmentation consists of adding training examples of non-contrast CT scans that are synthetically generated by CyTran. As a secondary contribution, we extend ViT-V-Net \cite{Chen-VITVN-2021} by employing multiple cascades at inference time, further improving the registration results by a considerable margin.

\subsection{Data Sets}

In recent years, the open-source access to large medical databases \cite{Heimann-TMI-2009, Kiryati-JI-2021, Malone-NI-2013, Oakden-AR-2020, Sivaswamy-ISBI-2014} has accelerated the development of deep learning methods in the medical imaging field.
The organizers of the CHAOS challenge \cite{Kavur-MIA-2021} released a medical data set containing CT and MRI data. The CT data was acquired from the upper abdomen area of 40 patients during the portal venous phase, after contrast agent injection. 
Moen et al.~\cite{Moen-MP-2021} developed a data set of CT scans from 299 patients for three types of clinical exams: non-contrast head CT scans, low-dose non-contrast chest scans and contrast-enhanced CT scans of the abdomen. Bilic et al.~\cite{Bilic-arXiv-2019} released a data set which consists of 140 CT scans, each having five organs labeled: lung, bones, liver, kidneys and bladder. The data set blends examples from a wide variety of sources, including abdominal and full-body, contrast and non-contrast, low-dose and high-dose CT scans. The data sets of Moen et al.~\cite{Moen-MP-2021} and Bilic et al.~\cite{Bilic-arXiv-2019} contain both contrast and non-contrast CT scans, but these are taken for different body sections. In contrast, our data set contains contrast and non-contrast CT scans of the same body section, the chest. To the best of our knowledge, Coltea-Lung-CT-100W is the first public data set formed entirely of triphasic lung CT scans, meaning that there are three 3D scans for each patient, corresponding to the native, early portal venous, and late arterial phases, respectively.

\section{Image Translation Method}

We propose a cycle-consistent generative adversarial transformer, which employs a generative visual transformer network to translate lung CT images between two different contrasts, e.g.~native, venous or arterial. Our approach is inspired by the success of cycle-consistent GANs \cite{Zhu-ICCV-2017} in image-to-image translation for style transfer. Based on the assumption that style is easier to transfer than other aspects, e.g.~geometrical deformations, cycle-GANs can replace the style of an image with a different style, while keeping its content. Our task involves style transfer between lung CT images acquired in different contrast phases. The contrast substance introduces HU changes for specific anatomical structures, such as tumors or blood vessels. However, the structures themselves should not exhibit geometrical changes between contrast phases, other than those caused by small movements of the patient, such as movements generated by respiration. While the changes between different contrast phases can be assimilated to style changes, we underline that the changes apply only to specific anatomical structures. Hence, to accurately mimic the contrast changes, the employed style transfer model should be capable of recognizing anatomical structures. We believe that models equipped with the power to extract and use global information have a higher capacity of recognizing anatomical structures. We thus conjecture that generative transformers can outperform convolutional generators at the task of learning to reproduce or unwind the changes caused by the contrast substance.

We propose a \textbf{cy}cle-consistent architecture based on generative adversarial \textbf{tran}sformers, termed \emph{CyTran}, to transfer CT scans between different contrast phases. Following the CycleGAN framework \cite{Zhu-ICCV-2017}, CyTran is formed of two discriminators and two generators. The neural architecture of the discriminators is identical to the architecture used by Zhu et al.~\cite{Zhu-ICCV-2017}. In a preliminary evaluation stage, we tried to replace the convolutional discriminators with transformers, but we observed that this change makes the discriminators too powerful with respect to the generative transformers. For this reason, we turned our attention to replacing the generative models only. We next describe in detail the proposed generative architecture as well as the entire optimization process.

\begin{figure*}[!t]
\begin{center}
\centerline{\includegraphics[width=1\linewidth]{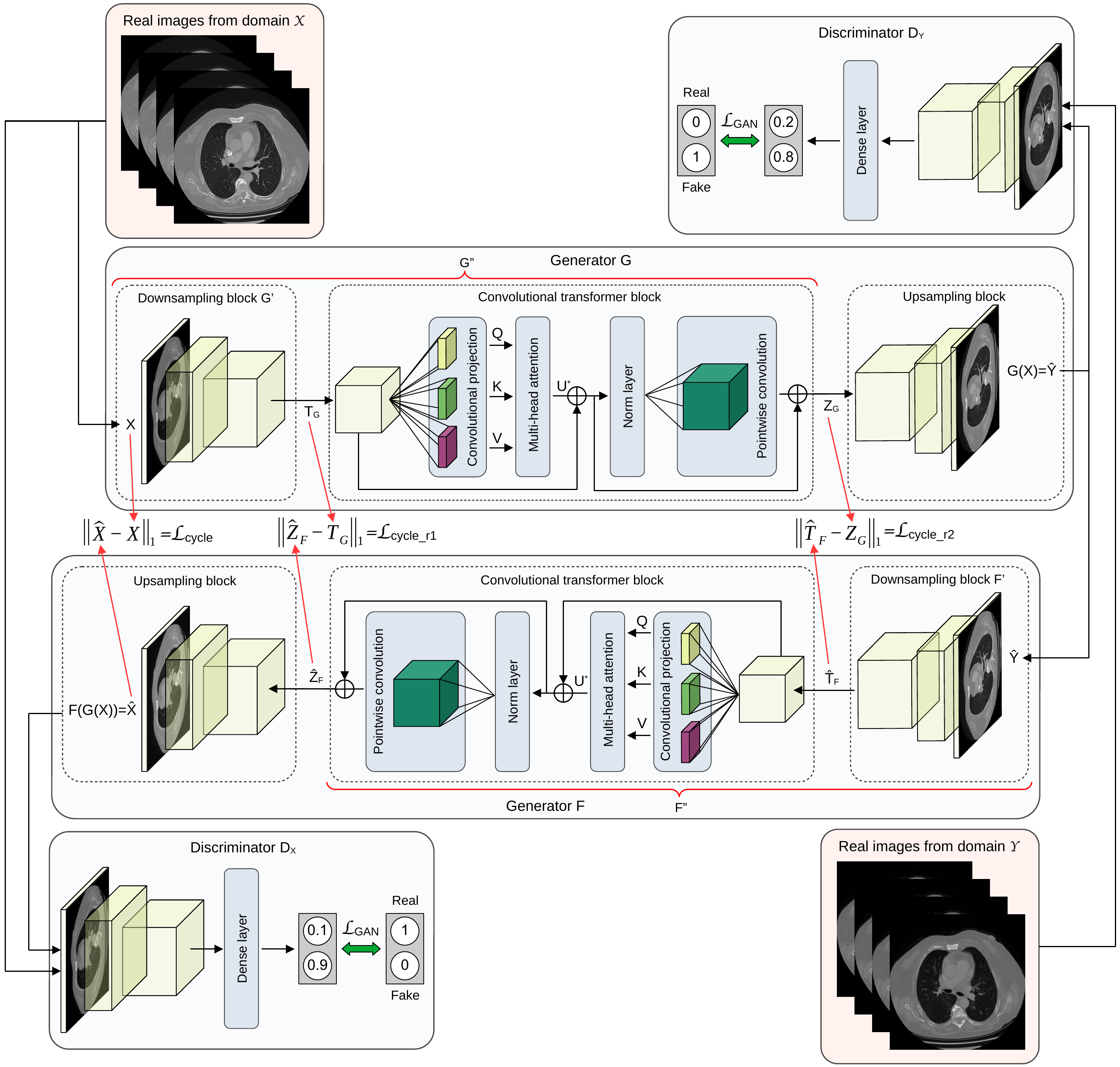}}
\caption{Our convolutional transformer for CT image translation from domain $\mathcal{X}$ to domain $\mathcal{Y}$, based on multi-level cycle consistency. Each generator is formed of a downsampling block comprising convolutional layers, a convolution transformer block comprising a multi-head self-attention mechanism, and an upsampling block comprising transposed convolutions. The source CT image $\boldsymbol{X}$ is translated using the generative transformer $G$ into the target CT image $\boldsymbol{\hat{Y}}$. The target CT image $\boldsymbol{\hat{Y}}$ is translated back to the original domain $\mathcal{X}$ through the generative transformer $F$. The generative transformer $G$ and the discriminator $D_Y$ are optimized in an adversarial fashion, just as in any other GAN. In addition, the model is optimized with respect to the cycle-consistency losses between the original source CT image $\boldsymbol{X}$ and the fake CT image $\boldsymbol{\hat{X}}$, as well as between the intermediate representations $\boldsymbol{T}_G$ and $\boldsymbol{\hat{Z}}_F$ on the one hand, and $\boldsymbol{Z}_G$ and $\boldsymbol{\hat{T}}_F$ on the other. Analogous steps are carried out for translating from domain $\mathcal{Y}$ to domain $\mathcal{X}$ (not represented in the image to improve clarity and ease readability). Best viewed in color.}
\label{fig_cytran}
\end{center}
\end{figure*}


\subsection{Generative Convolutional Transformer Architecture}

As we aim to benefit from the modeling power of transformers while being able to generate high-resolution CT images, we design a generative convolutional transformer with a manageable number of parameters. As illustrated in Figure \ref{fig_cytran}, our generator is formed of a convolutional downsampling block, a convolutional transformer block, and a deconvolutional upsampling block. We underline that, without the convolutional downsampling block and the replacement of dense layers with convolutional layers inside the transformer block, the transformer would not be able to learn to generate images larger than $64\times 64$ pixels, due to memory overflow (measured on an Nvidia GeForce RTX 3090 GPU with 24GB of VRAM). In contrast, we need a model capable of generating CT slices of $512\times512$ pixels. At this input resolution, our design changes significantly reduce the number of learnable parameters from 258 millions to 3.5 millions.

\subsection{Downsampling Block}

The downsampling block starts with a convolutional layer formed of $32$ filters with a spatial support of $7 \times 7$, which are applied using a padding of $3$ pixels to preserve the spatial dimension, while enriching the number of feature maps to $32$. Next, we apply three convolutional layers composed of $32$, $64$ and $128$ filters, respectively. All convolutional filters have a spatial support of $3 \times 3$ and are applied at a stride of $2$, using a padding of $1$. Each layer is followed by batch-norm \cite{Ioffe-ICML-2015} and Rectified Linear Units (ReLU) \cite{Nair-ICML-2010}. Let $\boldsymbol{T} \in \mathbb{R}^{h \times w \times c}$ denote the output tensor of the downsampling block. For an input CT slice of $512 \times 512$ pixels, the dimensions of $\boldsymbol{T}$ are $64 \times 64 \times 128$.

\subsection{Convolutional Transformer Block}

The downsampling block is followed by the convolutional transformer block, which provides an output tensor of the same size as the input tensor. Our convolutional transformer block is inspired by the block proposed by Wu et al.~\cite{Wu-arXiv-2021}.

The input tensor $\boldsymbol{T}$ is interpreted as a set of $h \cdot w$ overlapping visual tokens. In our implementation, we have $64\cdot 64 = 4096$ tokens, where each token is a tensor of $3 \times 3 \times 128$ components. The spatial dimensions of the visual tokens are determined by the receptive field of the filters in the next convolutional layer.

\noindent
{\bf Convolution projection.} 
In a vanilla transformer, a sequence of tokens is typically projected onto a set of weight matrices to obtain the queries $\boldsymbol{Q}$, the keys $\boldsymbol{K}$, and the values $\boldsymbol{V}$. Following Wu et al.~\cite{Wu-arXiv-2021}, the projections typically implemented through matrix multiplication are replaced by depthwise separable convolution operations, referred to as \emph{convolutional projection}. The convolutional projection is formed of three nearly identical projection blocks, with separate parameters. Each projection is a depthwise separable convolution block
\cite{Chollet-CVPR-2017} formed of two convolutional layers and a batch-normalization layer in between. 

The first layer in a projection block is a depthwise convolution with $128$ filters, each having a receptive field of $3 \times 3$. The projection block producing the queries is configured with a stride of $1$ (generating activation maps of $64 \times 64$), while the other projection blocks use a stride of $2$ (generating activation maps of $32\times32$). The padding is $1$ for all three blocks. The output passes through a batch-norm, before going into the third layer. The third layer applies pointwise convolution with $64$ filters. We hereby note that, in pointwise convolution, the filters always have a spatial support of $1 \times 1$ and are applied at a stride of $1$, without padding. Finally, the output tensors are reshaped into matrices by flattening the activation maps, while preserving the number of channels.

Let $\boldsymbol{W}\!_Q$, $\boldsymbol{W}\!_K$ and $\boldsymbol{W}\!_V$ denote the learnable parameters of the three projection blocks. The query, key and value embeddings are computed as follows:
\begin{equation}
\begin{split}
\boldsymbol{Q} &= \mbox{conv\_projection}\left(\boldsymbol{T}, \boldsymbol{W}\!_Q \right),\\
\boldsymbol{V} &= \mbox{conv\_projection}\left(\boldsymbol{T}, \boldsymbol{W}\!_V \right),\\
\boldsymbol{K} &= \mbox{conv\_projection}\left(\boldsymbol{T}, \boldsymbol{W}\!_K \right),\\
\end{split}
\end{equation}
where $\boldsymbol{Q} \in \mathbb{R}^{n_q \times d_q}$, $\boldsymbol{K} \in \mathbb{R}^{n_k \times d_k}$ and $\boldsymbol{V} \in \mathbb{R}^{n_v \times d_v}$. For the subsequent operation involving matrix multiplications, we need $d_q=d_k$, and $n_k=n_v$. In our implementation, $n_q=4096$ (obtained by flattening $64 \times 64$ activation maps) and $n_k=n_v=1024$ (obtained by flattening $32 \times 32$ activation maps). Due to the equal number of filters in the pointwise convolution in all three blocks, $d_q=d_k=d_v=64$.

We underline that the goal of adding the convolutional projection is to achieve additional modeling power of the local spatial context from the output of the subsequent multi-head attention layer.


\noindent
{\bf Multi-head self-attention.}
The convolutional projection layer is followed by a multi-head self-attention mechanism. The goal of self-attention is to capture the interaction among all tokens by encoding each entity in terms of the global contextual information.
Given a sequence of items, the self-attention mechanism estimates the relevance of an item to other items, e.g.~which visual token embeddings are likely to come together in a tensor. 
Basically, the self-attention layer updates each visual token by aggregating global information from the complete input tensor.

The output $\boldsymbol{U} \in \mathbb{R}^{n_q \times d_v}$ of the self-attention layer is given by:
\begin{equation}
\boldsymbol{U} = \mbox{softmax}\left( \frac{\boldsymbol{Q\cdot K^{\top}}}{\sqrt{d_k}} \right) \cdot \boldsymbol{V},
\end{equation}
where $\boldsymbol{K^{\top}}$ is the transpose of $\boldsymbol{K}$. For a given token, the self-attention computes the dot products of the query with all keys, which are then normalized using the softmax operator to get the attention scores.  According to Vaswani et al.~\cite{Vaswani-NIPS-2017}, the magnitudes of the dot products grow proportionally to $d_k$, propelling softmax into regions where it has extremely small gradients. To restore the gradient magnitudes, the dot products are scaled with respect to $d_k$. Each entity then becomes the weighted sum of all entities in the sequence, where the weights are given by the attention scores. At this point, $\boldsymbol{U}$ is a matrix of $4096 \times 64$ components. Upon reshaping the $4096$-dimensional vectors back into activation maps, we obtain a tensor of $64 \times 64 \times 64$ components.
In order to encapsulate multiple complex relationships among different tokens in the tensor $\boldsymbol{T}$, we employ a multi-head attention module \cite{Vaswani-NIPS-2017}. Each head $i \in  \{1,..., n_h\}$ comprises a convolution projection and a self-attention mechanism, having a particular set of learnable parameters $\{\boldsymbol{W}\!_{Q_i}, \boldsymbol{W}\!_{K_i}, \boldsymbol{W}\!_{V_i}\}$, where $n_h$ is the number of heads. Following the configuration for the CvT blocks used in stage 3 by Wu et al.~\cite{Wu-arXiv-2021}, we set $n_h=6$ in our block.

To form the output of the entire multi-head attention module, we concatenate the output tensors in the channel dimension, obtaining a tensor of $64 \times 64 \times 384$ components. A pointwise convolution with $128$ filters brings the dimension of the output tensor down to $64 \times 64 \times 128$ components. Let $\boldsymbol{U}^*$ denote the final output tensor of the multi-head self-attention module. We underline that the dimension of $\boldsymbol{U}^*$ coincides with the dimension of the input tensor $\boldsymbol{T}$, i.e.~$\boldsymbol{U}^* \in \mathbb{R}^{h \times w \times c}$.

\noindent
{\bf Pointwise convolution.}
After the multi-head attention layer, the output is summed up with the input of the convolutional projection and fed into a batch-normalization layer. Unlike the vast majority of transformers, we introduce a pointwise convolutional block instead of a multi-layer perceptron as the last processing step of the transformer block, further reducing the number of learnable parameters. Our convolutional block is formed of two consecutive pointwise convolutional layers, the first one being formed of $512$ filters and the second one being formed of $128$ filters. We use Gaussian Error Linear Units (GELU) \cite{Hendrycks-arXiv-2016} after the first pointwise convolutional layer.

Next, the input of the norm layer is added to the output of our pointwise convolutional block, resulting in the final output of our convolutional transformer block, denoted as $\boldsymbol{Z}$. 

\subsection{Upsampling Block}

The last block of our convolutional transformer applies upsampling operations, being designed to revert the transformation of the downsampling block. The upsampling block is formed of three transposed convolutional layers comprising $128$, $64$ and $32$ filters, respectively. All kernels have a spatial support of $3 \times 3$, being applied at a stride of $2$, using a padding of $1$. As for the downsampling block, we introduce batch-norm and ReLU activations after each transposed convolutional layer. Finally, we employ a convolutional layer with one filter to reduce the number of channels from $32$ back to $1$. The size of the receptive field of this final filter is $7 \times 7$. We use a padding of $3$ to preserve the spatial dimensions, obtaining an output image of $512 \times 512$ pixels.

\subsection{Learning on Unpaired CT Slices}


We use the following notations throughout the remainder of this work. Let $(\mathcal{X}, \mathcal{Y})$ denote the pair of source and target domains. Since we are interested in translating contrast CT scans to non-contrast CT scans and vice versa, the pair $(\mathcal{X}, \mathcal{Y})$ can take one of the following values: (native, arterial), (native, venous), (venous, native), (arterial, native). For our application purposes, we are not interested in translating the (arterial, venous) and (venous, arterial) pairs. Let $\boldsymbol{X}$ denote a sample from domain $\mathcal{X}$ and $\boldsymbol{Y}$ a sample from domain $\mathcal{Y}$, respectively.

In Figure~\ref{fig_cytran}, we illustrate the generative process based on cycle-consistency for the CT image $\boldsymbol{X}$. The source CT image $\boldsymbol{X}$ is translated using the generative transformer $G$ into $\boldsymbol{\hat{Y}}$, to make it seem that $\boldsymbol{\hat{Y}}$ belongs to the target domain $\mathcal{Y}$. The target CT image $\boldsymbol{\hat{Y}}$ is translated back to the original domain $\mathcal{X}$ through the generative transformer $F$. The generative transformer $G$ is optimized to fool the discriminator $D_Y$, while the discriminator $D_Y$ is optimized to separate synthesized CT images from real samples, in an adversarial fashion. In addition, the network is optimized with respect to the reconstruction error computed between the original sample $\boldsymbol{X}$ and the cycle-generated CT sample $\boldsymbol{\hat{X}}$. Adding the reconstruction error to the overall loss function ensures the cycle-consistency for domain $\mathcal{X}$. Moreover, we introduce additional losses to ensure the cycle-consistency at multiple representation levels. More precisely, we propose to add a cycle-consistency loss between the intermediate representations $\boldsymbol{T}_G$ and $\boldsymbol{\hat{Z}}_F$, and another cycle-consistency loss between the intermediate representations $\boldsymbol{Z}_G$ and $\boldsymbol{\hat{T}}_F$, respectively.

An analogous training process is carried out to ensure cycle-consistency for domain $\mathcal{Y}$. The complete loss function to optimize CyTran for contrast translation in both directions is:
\begin{equation}\label{eq_cytran}
\begin{split}
\!\!\!\mathcal{L}_{CyTran} &(G,F,D_X,D_Y,\boldsymbol{X},\boldsymbol{Y})=\mathcal{L}_{GAN} (G,D_Y,\boldsymbol{X},\boldsymbol{Y})\\
& +\mathcal{L}_{GAN} (F,D_X,\boldsymbol{X},\boldsymbol{Y})+ \lambda \cdot \mathcal{L}_{cycle} (G,F,\boldsymbol{X},\boldsymbol{Y})\\
& + \beta \cdot \left( \mathcal{L}_{cycle\_r1} (G,F,\boldsymbol{X},\boldsymbol{Y}) + \mathcal{L}_{cycle\_r2} (G,F,\boldsymbol{X},\boldsymbol{Y})  \right),
\end{split}
\end{equation}
where, $G$ and $F$ are generative transformers, $D_X$ and $D_Y$ are convolutional discriminators, $\boldsymbol{X}$ is a CT slice from contrast phase $\mathcal{X}$, and $\boldsymbol{Y}$ is a CT slice from contrast phase $\mathcal{Y}$. The hyperparameters $\lambda$ and $\beta$ control the importance of the cycle-consistency losses with respect to the two GAN losses. The first GAN loss is the least squares loss that corresponds to the translation from domain $\mathcal{X}$ to domain $\mathcal{Y}$:
\begin{equation}
\begin{split}
\!\!\!\mathcal{L}_{GAN} (G,D_Y,\boldsymbol{X},\boldsymbol{Y})&=E_{\boldsymbol{Y} \sim P_{data}(\boldsymbol{Y})} \left[(D_Y (\boldsymbol{Y}))^2\right]\\
&+E_{\boldsymbol{X} \sim P_{data} (\boldsymbol{X})} \left[(1\!-\!D_Y (G(\boldsymbol{X})))^2\right]\!,
\end{split}
\end{equation}
where $E[\cdot]$ is the expected value and $P_{data}(\cdot)$ is the probability distribution of data samples. Analogously, the second GAN loss is the least squares loss that corresponds to the translation from domain $\mathcal{Y}$ to domain $\mathcal{X}$:
\begin{equation}
\begin{split}
\!\!\!\mathcal{L}_{GAN} (F,D_X,\boldsymbol{X},\boldsymbol{Y})&=E_{\boldsymbol{X} \sim P_{data}(\boldsymbol{X})} \left[(D_X (\boldsymbol{X}))^2\right]\\
&+E_{\boldsymbol{Y} \sim P_{data} (\boldsymbol{Y})} \left[(1\!-\!D_X (F(\boldsymbol{Y})))^2\right]\!.
\end{split}
\end{equation}

The cycle-consistency loss applied in Equation~\eqref{eq_cytran} between the inputs and the outputs of the generators $G$ and $F$ is defined as the sum of cycle-consistency losses for both translations:
\begin{equation}
\begin{split}
\mathcal{L}_{cycle} (G,F,\boldsymbol{X},\boldsymbol{Y})&=E_{\boldsymbol{X} \sim P_{data} (\boldsymbol{X})} \left[ \left\lVert F(G(\boldsymbol{X}))-\boldsymbol{X} \right\rVert_1 \right]\\
&+E_{\boldsymbol{Y} \sim P_{data}(\boldsymbol{Y})} \left[ \left\lVert G(F(\boldsymbol{Y}))-\boldsymbol{Y}\right\rVert_1 \right], 
\end{split}
\end{equation}
where $\left\lVert \cdot \right\rVert_1$ is the $l_1$ norm. 

In Equation~\eqref{eq_cytran}, we introduce two cycle-consistency losses for the intermediate representations of the generators, but before defining the losses, we need to introduce the required notations. Let $\boldsymbol{T}_G$ and $\boldsymbol{Z}_G$ represent the features before and after the transformer block in $G$, and $\boldsymbol{T}_F$ and $\boldsymbol{Z}_F$ represent the features before and after the transformer block in $F$. Let $G'$ and $F'$ denote the downsampling blocks of the generators $G$ and $F$, respectively. With these notations, $\boldsymbol{T}_G = G'(\boldsymbol{X})$, $\boldsymbol{T}_F = F'(\boldsymbol{Y})$, $\boldsymbol{\hat{T}}_G = G'(\boldsymbol{\hat{X}})=G'(F(Y))$ and $\boldsymbol{\hat{T}}_F=F'(\boldsymbol{\hat{Y}})=F'(G(\boldsymbol{X}))$. Similarly, let $G''$ and $F''$ denote the sub-networks formed of the downsampling block and the convolutional transformer of the generators $G$ and $F$, respectively. With these notations, $\boldsymbol{Z}_G = G''(\boldsymbol{X})$, $\boldsymbol{Z}_F = F''(\boldsymbol{Y})$, $\boldsymbol{\hat{Z}}_G=G''(\boldsymbol{\hat{X}})=G''(F(\boldsymbol{Y}))$ and $\boldsymbol{\hat{Z}}_F=F''(\boldsymbol{\hat{Y}})=F''(G(\boldsymbol{X}))$. Now, we define the cycle-consistency loss for the intermediate representations $\boldsymbol{T}_G$ and $\boldsymbol{T}_F$ before the convolutional transformer block as follows:
\begin{equation}
\begin{split}
\mathcal{L}_{cycle\_r1} (G,F,\boldsymbol{X},\boldsymbol{Y})&=E_{\boldsymbol{T}_G \sim P_{data} (G'(\boldsymbol{X}))} \left[ \left\lVert F''(G(\boldsymbol{X}))-G'(\boldsymbol{X}) \right\rVert_1 \right]\\
&+E_{\boldsymbol{T}_F \sim P_{data} (F'(\boldsymbol{Y}))} \left[ \left\lVert G''(F(\boldsymbol{Y}))-F'(\boldsymbol{Y})\right\rVert_1 \right]\\
&=E_{\boldsymbol{T}_G \sim P_{data} (G'(\boldsymbol{X}))} \left[ \left\lVert \boldsymbol{\hat{Z}}_F-\boldsymbol{T}_G \right\rVert_1 \right]\\
&+E_{\boldsymbol{T}_F \sim P_{data} (F'(\boldsymbol{Y}))} \left[ \left\lVert \boldsymbol{\hat{Z}}_G-\boldsymbol{T}_F\right\rVert_1 \right].
\end{split}
\end{equation}
Analogously, we define the cycle-consistency loss for the intermediate representations $\boldsymbol{Z}_G$ and $\boldsymbol{Z}_F$ after the convolutional transformer block as follows:
\begin{equation}
\begin{split}
\mathcal{L}_{cycle\_r2} (G,F,\boldsymbol{X},\boldsymbol{Y})&=E_{\boldsymbol{Z}_G \sim P_{data} (G''(\boldsymbol{X}))} \left[ \left\lVert F'(G(\boldsymbol{X}))-G''(\boldsymbol{X}) \right\rVert_1 \right]\\
&+E_{\boldsymbol{Z}_F \sim P_{data}(F''(\boldsymbol{Y}))} \left[ \left\lVert G'(F(\boldsymbol{Y}))-F''(\boldsymbol{Y})\right\rVert_1 \right]\\
&=E_{\boldsymbol{Z}_G \sim P_{data} (G''(\boldsymbol{X}))} \left[ \left\lVert \boldsymbol{\hat{T}}_F-\boldsymbol{Z}_G \right\rVert_1 \right]\\
&+E_{\boldsymbol{Z}_F \sim P_{data}(F'(\boldsymbol{Y}))} \left[ \left\lVert \boldsymbol{\hat{T}}_G-\boldsymbol{Z}_F\right\rVert_1 \right].
\end{split}
\end{equation}

\section{Image Registration Method}

Unsupervised medical image registration can fail to properly align contrast CT scans to non-contrast CT scans, especially in the regions highlighted by the contrast agent, which are of primary interest. To alleviate such failure cases, we propose to employ CyTran on the contrast CT scans to eliminate differences induced by the contrast agent. We underline that switching the roles of contrast and non-contrast CT scans in image registration leads to an equivalent problem. For simplicity, we only study the alignment of contrast CT scans to non-contrast CT scans.

Let $\boldsymbol{X} \in \mathcal{X}$ denote a source 3D CT scan belonging to the contrast phase $\mathcal{X}$ (either venous or arterial), and $\boldsymbol{Y} \in \mathcal{Y}$ a target 3D CT scan belonging to the non-contrast phase $\mathcal{Y}$ (native). Prior to the alignment of voxels in $\boldsymbol{X}$ to $\boldsymbol{Y}$, we translate all slices in the CT scan $\boldsymbol{X}$ to the distribution $\mathcal{Y}$, obtaining $\boldsymbol{\hat{X}}$. Then, we align $\boldsymbol{\hat{X}}$ to $\boldsymbol{Y}$, both belonging to the same distribution $\mathcal{Y}$, and we obtain the displacement field $\boldsymbol{M_{\hat{X}}}$. Finally, we apply the displacement field $\boldsymbol{M_{\hat{X}}}$ to $\boldsymbol{X}$ in order to obtain the final alignment result.

To perform the alignment, we rely on the state-of-the-art ViT-V-Net model introduced by Chen et al.~\cite{Chen-VITVN-2021}. ViT-V-Net is a hybrid convolutional-transformer architecture for self-supervised volumetric medical image registration. The architecture contains three convolutional blocks and two max-pooling operations, which form the first part of the model, producing a high-level feature representation. The resulting feature maps are subsequently divided into patches, which are further projected into tokens by a projection layer. The tokens are appended with positional embeddings, then fed into a transformer-based encoder. The transformer-based encoder consists of $12$ transformer blocks comprising multi-head self-attention and dense layers, just as ViT \cite{Dosovitskiy-ICLR-2020}. Next, the output of the transformer-based encoder is reshaped and passed to a convolutional decoder. ViT-V-Net  also contains long skip connections between the encoder and the decoder. At the end, the output is decoded into a dense displacement field. The ViT-V-Net model is optimized towards minimizing the mean squared error between the moving image and the reference image. To ensure the smoothness of the displacement field, a diffusion regularizer is added to the optimization objective.


\begin{figure}[!t]
\begin{center}
\centering
\includegraphics[width=0.6\columnwidth]{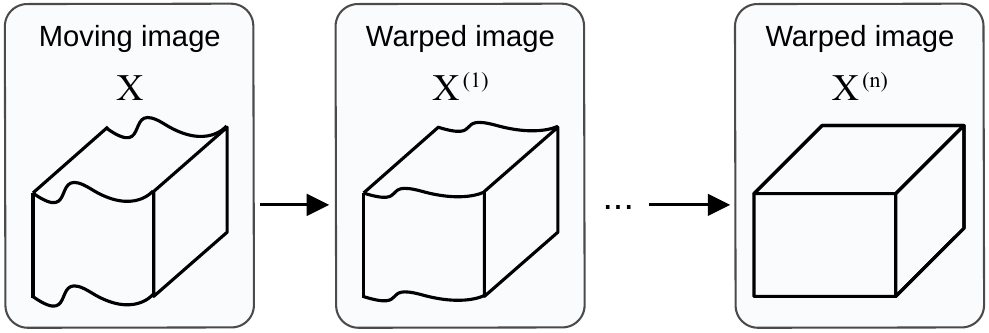}
\caption{Cascaded algorithm for volumetric image registration. The moving 3D CT scan $\boldsymbol{X}$ is registered in a recursive cascade based on $n$ steps. Thus, $\boldsymbol{X}^{(i)}$ is the warped image at step $i$, $\forall i \in \{1, ...,n \}$.}
\label{fig_cascade}
\end{center}
\end{figure}
 
As another contribution, we extend ViT-V-Net~\cite{Chen-VITVN-2021} to a cascaded registration algorithm, which is intuitively illustrated in Figure \ref{fig_cascade}. Let $R$ be the registration model, $\boldsymbol{X} \in \mathbb{R}^{h \times w \times d}$ the moving 3D image, and $\boldsymbol{X}^{(1)} \in \mathbb{R}^{h \times w \times d}$ the warped image, such that $R(\boldsymbol{X}) = \boldsymbol{X}^{(1)}$. We propose to apply multiple cascades at inference time, by passing the output several times through the model:
\begin{equation}
R\left(\boldsymbol{X}^{(i-1)}\right) = \boldsymbol{X}^{(i)}, \forall i \in \{1, ...,n \}.
\end{equation}

The recursive processing progressively reduces the alignment differences, leading to a superior result. This statement is supported by the experiments presented below.

\begin{table}
\centering
\noindent
\caption{The number of triphasic CT scans and individual slices from the Coltea-Lung-CT-100W data set.}
\setlength\tabcolsep{4.5pt}
\begin{tabular}{|l|ccc|c|}
\hline
       & Training  & Validation    & Test      & Total\\ 
\hline
\hline
\#scans     & 70        & 15            & 15        & 100 \\ 
\hline
\#images    & 25,311     & 5,937        & 6,042      & 37,290 \\ 
\hline
\end{tabular}
\label{tab_dataset}
\end{table}

\begin{figure*}[!t]
\begin{center}
\centering
{\includegraphics[width=0.85\linewidth]{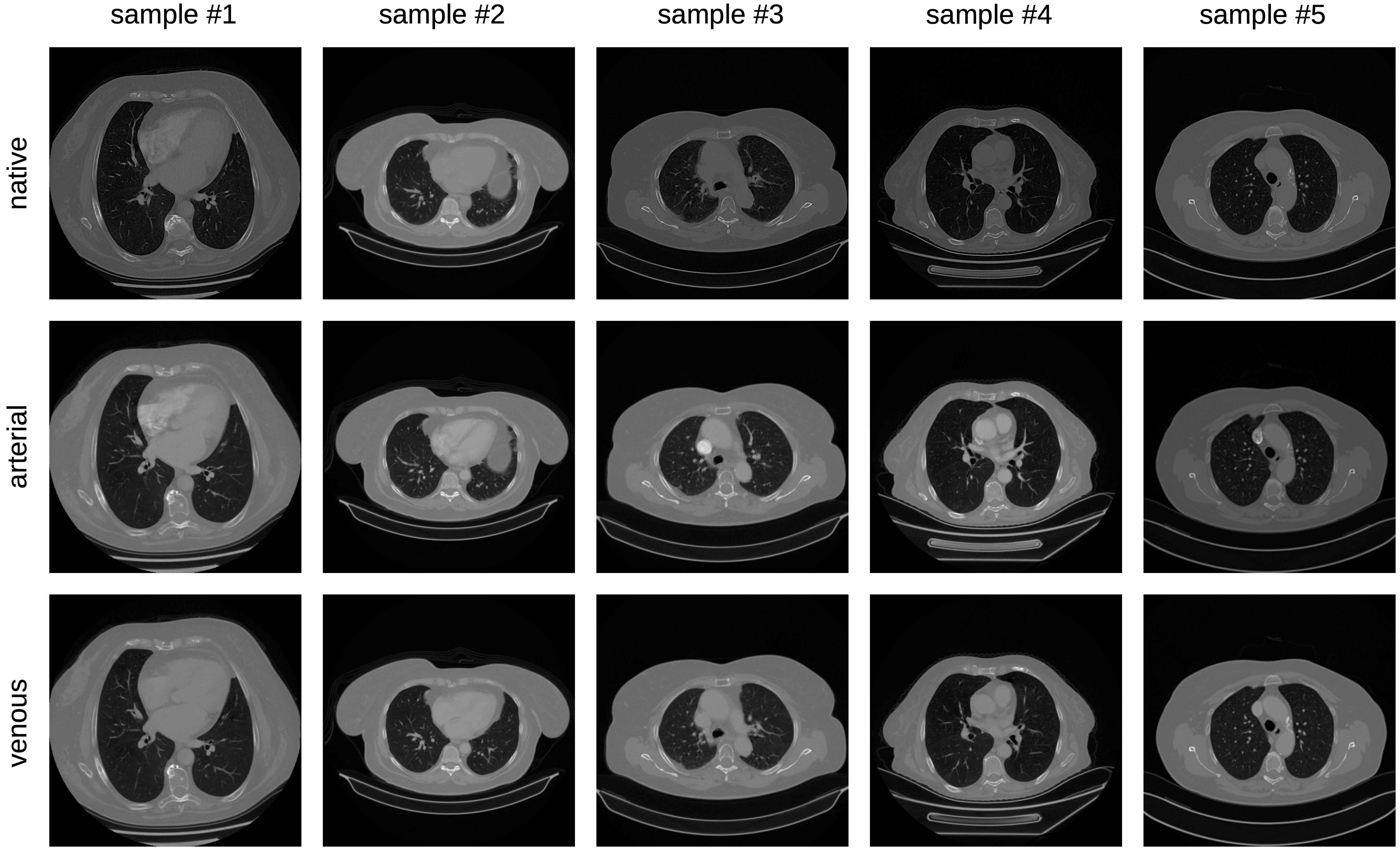}}
\caption{A set of randomly selected samples from the Coltea-Lung-CT-100W data set. For each of the five samples, we illustrate the native, arterial and venous phases, from top to bottom.}
\label{fig_data_samples}
\end{center}
\end{figure*}

\begin{figure*}[!t]
\begin{center}
\centering
{\includegraphics[width=0.45\linewidth]{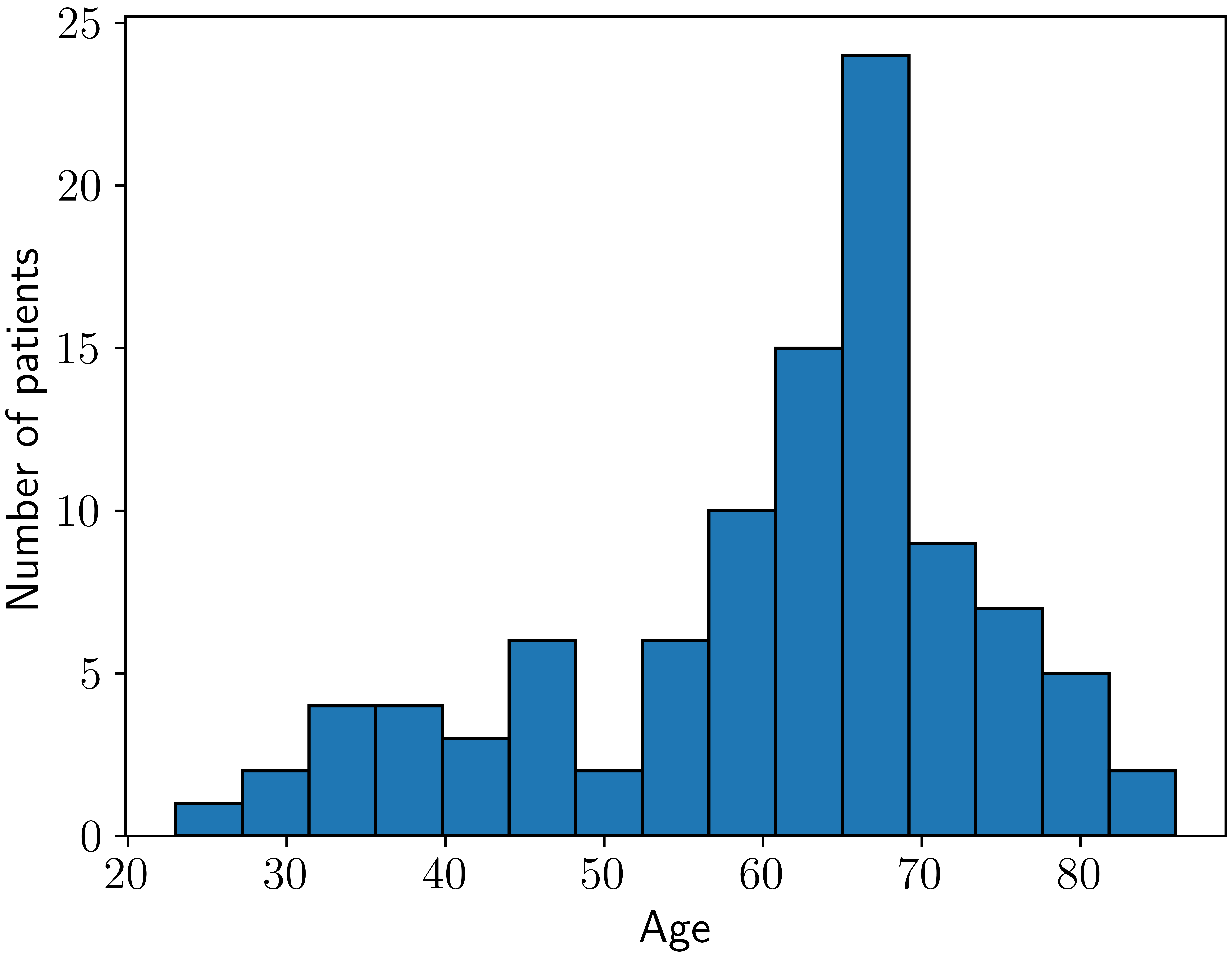}}
\caption{Age distribution of the 100 Romanian female patients who agreed to donate their anonymized CT scans for research purposes and inclusion in the Coltea-Lung-CT-100W data set.}
\label{fig_age_dist}
\end{center}
\end{figure*}

\section{Data Set}

We release a novel data set entitled \emph{Coltea-Lung-CT-100W}, which consists of $100$ triphasic lung CT scans. The scans are collected from $100$ female patients and represent the same body section. There is one triphasic CT scan per patient. The slices are selected having as anatomical landmarks the 7th cervical vertebra cranially and the 12th thoracic vertebra caudally. A triphasic scan is formed of a native (non-contrast) scan, an early portal venous scan, and a late arterial scan. In our data set, the three CT scans forming a triphasic scan always have the same number of slices. Hence, each triphasic scan comes roughly aligned, i.e.~the slice at index $i$ gathered during the native phase should correspond to the slice at index $i$ gathered during each contrast phase. This is illustrated in Figure~\ref{fig_data_samples}. The number of slices may differ from one patient to another. Indeed, the number of slices per scan ranges between $64$ and $229$, and the total number of slices is $37,290$. The size of a CT slice is $512 \times 512$ pixels and the slice thickness varies between $1.25$ and $3\; mm$. The resolution of a pixel is $1 \times 1\; mm^2$.

\begin{figure*}[!t]
\begin{center}
\centering
{\includegraphics[width=0.4\linewidth]{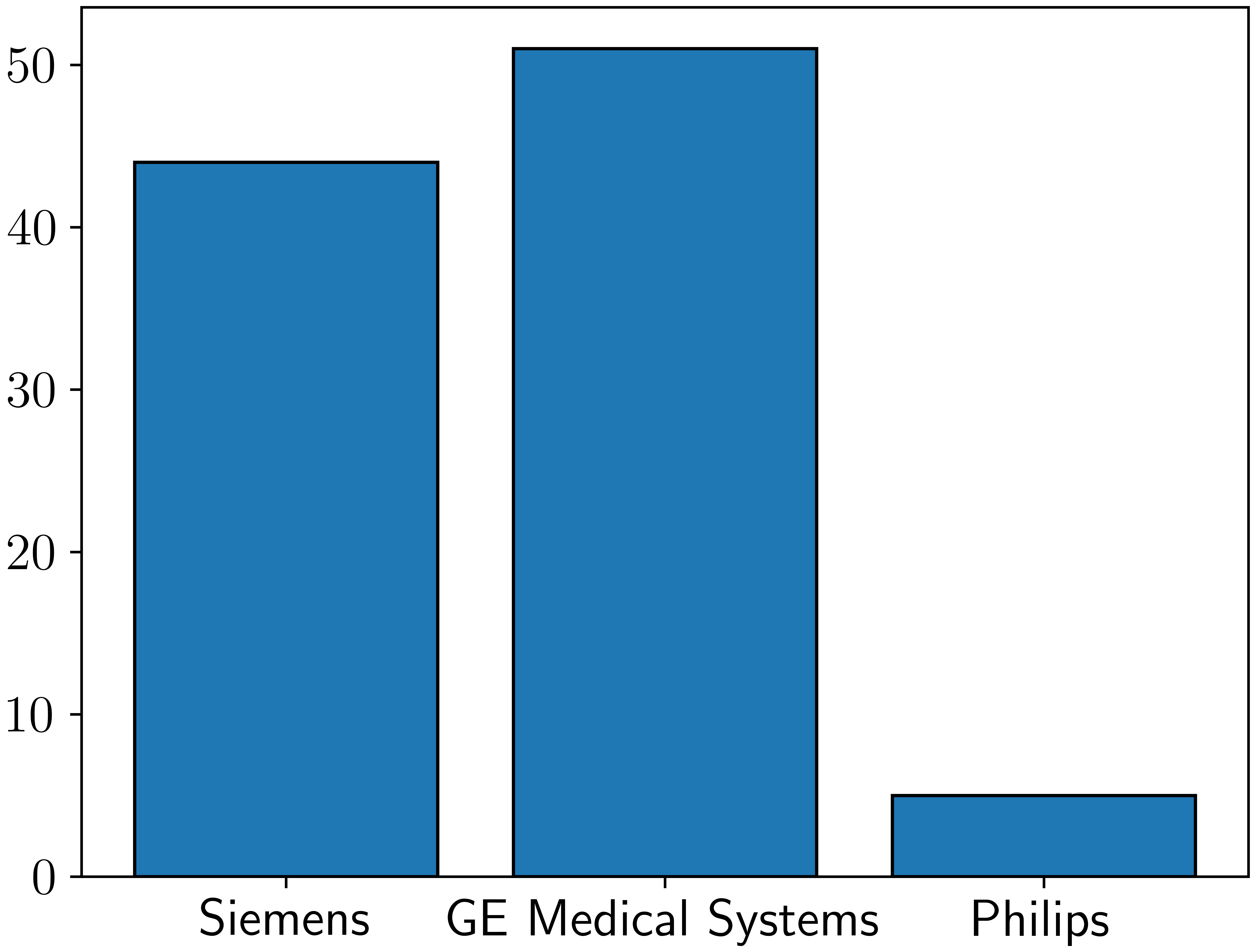}}
\caption{Manufacturer distribution of the CT scanners used to collect the CT scans from the Coltea-Lung-CT-100W data set.}
\label{fig_manu_dist}
\end{center}
\end{figure*}

The female patients are citizens of different parts of Romania and belong to different age groups, as illustrated through the bar chart presented in Figure \ref{fig_age_dist}. We refrain from providing more details about the patients to preserve their anonymity. The CT scans are produced by different CT scanners from three different manufacturers. The distribution of scans per manufacturer is shown in Figure \ref{fig_manu_dist}.

We split our data set into three subsets, one for training (70 scans), one for validation (15 scans), and one for testing (15 scans). We report the number of slices in each subset in Table~\ref{tab_dataset}. Our data set is stored as anonymized raw DICOM files and can be freely downloaded from \url{https://github.com/ristea/cycle-transformer}. To the best of our knowledge, Coltea-Lung-CT-100W is the only publicly available data set of triphasic lung CT scans.


\section{Experiments}

We study two tasks on Coltea-Lung-CT-100W: 
\begin{itemize}
\item style transfer between contrast and non-constrast CT slices, considering the following pairs of contrast phases: native$\rightarrow$venous, venous$\rightarrow$native, native$\rightarrow$arterial, arterial$\rightarrow$native.
\item volumetric image registration of contrast CT scans to non-contrast CT scans, considering the following pairs: venous$\rightarrow$native, arterial$\rightarrow$native.
\end{itemize}


\subsection{Style Transfer Experiments}

\subsubsection{Baselines}

We compare CyTran with five state-of-the-art style transfer methods. Since our data set is suitable for paired style transfer, we consider pix2pix \cite{Isola-CVPR-2017} as the first baseline. As CyTran is an approach capable of learning from unpaired images, we select four more baselines from the same category of methods, namely CycleGAN \cite{Zhu-ICCV-2017}, U-GAT-IT \cite{Kim-ICLR-2019}, CWT-GAN \cite{Lai-ICCV-2021}, and AttentionGAN \cite{Tang-TNNLS-2021}.
We note that GAN-based style transfer methods can introduce visual artifacts during the generation process, which could be problematic in medical practice. We thus consider important comparing the generative models with a \emph{no-transfer} baseline, which simply outputs the unprocessed input image. If this baseline outperforms a generative model, it indicates the respective model is unreliable, introducing too many artifacts.

\subsubsection{Performance evaluation}

We report the mean absolute error (MAE), root mean square error (RMSE) and structure-similarity index measure (SSIM) between the $i$-th translated image in a source scan and the slice at the same index $i$ in the corresponding target scan. The SSIM measures the capability of keeping original structures unchanged, while the MAE and RMSE measure the ability of transferring the desired style, e.g.~the ability of raising HU levels of a tumor in native$\rightarrow$venous style transfer.

Furthermore, we asked three medical experts (two radiotherapists and one oncologist with experience in radiology) to independently and blindly vote for the best among six style transfer methods: pix2pix, CycleGAN, U-GAT-IT, CWT-GAN, AttentionGAN and CyTran. The experts evaluated a total of $200$ cases (with 8 slices per case to be analyzed), randomly picked from the test set. We selected $50$ cases for each of the four contrast pairs. One case is composed of an input slice, 6 translated slices (produced by the generative models) and the corresponding target slice. All slices are provided in DICOM format and viewed with a specialized software, such that the experts can visualize and compare the slices under multiple window and level settings. The annotators were instructed to analyze the translated images in order to observe structural deformations, the correctness of contrast change with respect to the target image, and the occurrence of visual artifacts with respect to the input CT image. To obtain reliable annotations, we relied on the many years (between $2$ and $35$) of working experience of the medical experts in analyzing CT scans. During the annotation, we did not disclose the matching between models and translated images to the experts, i.e.~the votes are blind. With each presented case, the translated images were randomly shuffled, so it would be impossible for the experts to know which translation method produced a certain image. As evaluation measures, we report the number of votes and the corresponding percentage for each method.

\subsubsection{Data preprocessing}

To avoid working with high values and ensure the stable training of deep learning models, we shifted the raw voxel values to the HU scale by subtracting the intercept (stored in the DICOM metadata), and we divided the resulting values by $1000$. We apply this preprocessing for all the models, including the baselines.

\begin{table*}[t]
\setlength{\tabcolsep}{1.5pt}
\renewcommand{\arraystretch}{1.2}
\centering
\caption{Style transfer results for the native$\rightarrow$venous, venous$\rightarrow$native, native$\rightarrow$arterial and arterial$\rightarrow$native contrast pairs. Our model is compared with several state-of-the-art baselines on the test set, in terms of MAE, RMSE and SSIM. The symbol $\uparrow$ means higher values are better, while $\downarrow$ means lower values are better. There are three ablated versions of CyTran added to the comparison: no attention (the transformer block is removed), no pointwise (the pointwise convolutions are replaced with dense layers, as in standard transformer blocks), no MLCC (our multi-level cycle-consistency loss is replaced with a standard cycle-consistency loss). The best results are highlighted in bold.}
\label{tab_style}
 \begin{tabular}{|l|ccc|ccc|ccc|ccc|}
 \hline
\multirow{2}{*}{Method} &
\multicolumn{3}{c|}{native$\rightarrow$venous} &
\multicolumn{3}{c|}{venous$\rightarrow$native} & \multicolumn{3}{c|}{native$\rightarrow$arterial} & \multicolumn{3}{c|}{arterial$\rightarrow$native} \\
  & \rotatebox{45}{MAE$\downarrow$} & \rotatebox{45}{RMSE$\downarrow$} & \rotatebox{45}{SSIM$\uparrow$} & \rotatebox{45}{MAE$\downarrow$} & \rotatebox{45}{RMSE$\downarrow$} & \rotatebox{45}{SSIM$\uparrow$} & \rotatebox{45}{MAE$\downarrow$} & \rotatebox{45}{RMSE$\downarrow$} & \rotatebox{45}{SSIM$\uparrow$} & \rotatebox{45}{MAE$\downarrow$} & \rotatebox{45}{RMSE$\downarrow$} & \rotatebox{45}{SSIM$\uparrow$}\\
 \hline
 \hline
 no-transfer                    & 0.072 & 0.160 & 0.656  & 0.072 & 0.160 & 0.656 & 0.072 & 0.163 & 0.664 & 0.072 & 0.163 & 0.664 \\ 
 \hline
 pix2pix \citep{Isola-CVPR-2017} & 0.070 & 0.165 & 0.729  & 0.076 & 0.180 & 0.646 & 0.064 & 0.157 & 0.738 & 0.075 & 0.174 & 0.648 \\ 
\hline
 U-GAT-IT \citep{Kim-ICLR-2019}  & 0.066 & 0.150 & 0.720  & 0.074 & 0.162 & 0.642 & 0.066 & 0.152 & 0.734 & 0.073 & 0.160 & 0.651 \\ 
\hline
 CycleGAN \citep{Zhu-ICCV-2017}  & 0.066 & 0.150 & 0.724  & 0.071 & 0.160 & 0.660 & 0.065 & 0.154 & 0.729 & 0.072 & 0.160 & 0.662  \\ 
\hline
 CWT-GAN \citep{Lai-ICCV-2021}  & 0.069 & 0.153 & 0.726  & 0.072 & 0.159 & 0.656 & 0.065 & 0.153 & 0.733 & 0.071 & 0.163 & 0.663  \\ 
\hline
 AttentionGAN \citep{Tang-TNNLS-2021}  & 0.064 & 0.150 & 0.730  & 0.072 & 0.159 & 0.659 & 0.064 & 0.151 & 0.737 & 0.073 & 0.159 & 0.659  \\ 
\hline
  CyTran (no attention)                  & 0.067 & 0.156 & 0.725 & 0.073 & 0.166 & 0.658 & 0.066 & 0.156 & 0.739 & 0.074 & 0.163 & 0.655 \\
   CyTran (no pointwise)                & 0.066 & 0.151 & 0.733 & 0.073 & 0.164 & 0.660 & 0.065 & 0.150 & 0.741 & 0.072 & 0.159 & 0.662 \\
 CyTran (no MLCC)                  & 0.063 & 0.147 & 0.739 & \textbf{0.070} & \textbf{0.157} & \textbf{0.664} & 0.063 & 0.149 & 0.742 & 0.070 & \textbf{0.156} & \textbf{0.668} \\ 
 CyTran (proposed)                  & \textbf{0.061} & \textbf{0.144} & \textbf{0.745} & \textbf{0.070} & \textbf{0.157} & \textbf{0.664} & \textbf{0.059} & \textbf{0.147} & \textbf{0.758} & \textbf{0.069} & \textbf{0.156} & \textbf{0.668} \\
 \hline
\end{tabular}
\end{table*}

\subsubsection{Hyperparameter tuning}

We train all generative models from scratch using Adam \cite{Kingma-ICLR-2014} for $70$ epochs, on mini-batches of two examples. For any additional hyperparameters of the baseline methods, we consider the default values proposed by the authors introducing the respective models \cite{Zhu-ICCV-2017,Isola-CVPR-2017,Kim-ICLR-2019, Lai-ICCV-2021, Tang-TNNLS-2021}, along with additional configurations (comprising various learning rates and loss weights) obtained through grid search. As we did not observe significant improvements when changing the hyperparameters for any of the baseline models, we decided to keep the default hyperparameter values as indicated by the corresponding authors \cite{Zhu-ICCV-2017,Isola-CVPR-2017,Kim-ICLR-2019, Lai-ICCV-2021, Tang-TNNLS-2021}. For our approach, we set the learning rate to $10^{-4}$, keeping the default values for the other parameters of Adam. We set the weights that control the importance of the cycle-consistency terms in Equation~\eqref{eq_cytran} to $\lambda=10$ and $\beta=1$. Following Zhu et al.~\cite{Zhu-ICCV-2017}, the same value of $\lambda$ is used for CycleGAN. We provide the code to reproduce our results at: \url{https://github.com/ristea/cycle-transformer}.

\subsubsection{Quantitative results}

We conducted style transfer experiments on four contrast pairs, comparing our approach against five state-of-the-art methods and the \emph{no-transfer} baseline. The corresponding results are shown in Table~\ref{tab_style}. First of all, we observe that CyTran is the only approach that consistently surpasses the \emph{no-transfer} baseline across all contrast pairs and evaluation metrics. For instance, our approach is the only one able to surpass the \emph{no-transfer} baseline for the arterial$\rightarrow$native transfer. Although pix2pix \cite{Isola-CVPR-2017} can leverage the paired nature of the data set, it seems to produce the lowest performance levels, being surpassed by CycleGAN \cite{Zhu-ICCV-2017}, U-GAT-IT \cite{Kim-ICLR-2019}, CWT-GAN, AttentionGAN \citep{Tang-TNNLS-2021} and CyTran. All in all, our method attains the highest performance levels in each and every experiment, always outperforming the baselines. This conclusion supports our conjecture that CyTran is a more suitable method for style transfer between contrast and non-contrast CT images.

\subsubsection{Ablation study}

We perform an ablation study to determine the impact of each component proposed in this work. The ablation results are reported in Table~\ref{tab_style}. In our first ablation experiment, we remove the transformer block, essentially transforming CyTran into a convolutional network without attention. This ablated configuration results in suboptimal results, being surpassed by most of the baselines. Hence, the first ablation experiment reveals the importance of adding the transformer block inside CyTran. In the second ablation experiment, we replace the pointwise convolutions with dense layers, as in standard transformer blocks. The resulting architecture obtains scores that are comparable to the baselines. In the third ablation experiment, we replace our multi-level cycle-consistency (MLCC) loss with a standard cycle-consistency loss. This ablated version of CyTran surpasses all state-of-the-art models. Moreover, when transferring images to the native domain, the performance level without MLCC is comparable to that of the full CyTran framework. This seems to indicate that the multi-level cycle-consistency loss is mostly useful when transferring non-contrast (native) slices to contrast (arterial or venous) slices. In summary, all our novel components are beneficial to the proposed model.

\begin{table*}[t]
\setlength{\tabcolsep}{1.9pt}
\renewcommand{\arraystretch}{1.2}
\centering
\caption{Subjective human evaluation results based on 50 cases randomly selected from the test set for each of the following contrast pairs: native$\rightarrow$venous, venous$\rightarrow$native, native$\rightarrow$arterial and arterial$\rightarrow$native (there are 200 cases in total, with 6 slices per case). The reported numbers represent blind votes awarded by three medical experts for each generative model.}
\label{tab_doctors}
 \begin{tabular}{|l|cccccc|cccccc|cccccc|}
 \hline
\multirow{5}{*}{Contrast pair} & 
\multicolumn{6}{c|}{Expert \#1} & 
\multicolumn{6}{c|}{Expert \#2} &
\multicolumn{6}{c|}{Expert \#3} \\
  & {\rotatebox{90}{pix2pix}} & {\rotatebox{90}{U-GAT-IT}} & {\rotatebox{90}{CycleGAN}} & {\rotatebox{90}{CWT-GAN}} & {\rotatebox{90}{AttentionGAN}} & {\rotatebox{90}{CyTran (ours)}} & {\rotatebox{90}{pix2pix}} &  {\rotatebox{90}{U-GAT-IT}} & {\rotatebox{90}{CycleGAN}} & {\rotatebox{90}{CWT-GAN}} & {\rotatebox{90}{AttentionGAN}} & {\rotatebox{90}{CyTran (ours)}} & {\rotatebox{90}{pix2pix}} &  {\rotatebox{90}{U-GAT-IT}} & {\rotatebox{90}{CycleGAN}} & {\rotatebox{90}{CWT-GAN}} & {\rotatebox{90}{AttentionGAN}} & {\rotatebox{90}{CyTran (ours)}}\\
   
\hline
\hline
native$\rightarrow$venous & 4 & 6 & 4 & 0 & 3 & 33 & 0 & 0 & 0 & 1 & 4 & 45 & 0 & 0 & 0 & 0 & 2 & 48 \\ 

\hline
venous$\rightarrow$native& 3 & 3 & 2 & 6 & 0 & 36 & 0 & 2 & 7 & 0 & 6 & 35 & 0 & 0 & 8 & 0 & 3 & 39\\

\hline
native$\rightarrow$arterial & 0 & 2 & 2 & 1 & 4 & 41 & 0 & 0 & 0 & 0 & 5 & 45 & 0 & 0 & 0 & 0 & 4 & 46\\ 

\hline
arterial$\rightarrow$native & 0 & 3 & 3 & 0 & 4 & 40 & 0 & 0 & 6 & 0 & 4 & 40 & 0 & 0 & 3 & 0 & 4 & 43\\ 

\hline
\hline
Overall votes & 7 & 14 & 11 & 7 & 11 & 150 & 0 & 2 & 13 & 1 & 19 & 165 & 0 & 0 & 11 & 0 & 13 & 176\\ 
\hline
Overall (\%) & 3.5\% & 7\% & 5.5\% & 3.5\% & 5.5\% & 75\% & 0\% & 1\% & 6.5\% & 0.5\% & 9.5\% & 82.5\% & 0\% & 0\% & 5.5\% & 0\% & 6.5\% & 88\%\\ 
\hline
\end{tabular}
\end{table*}

\subsubsection{Subjective human evaluation results}


While the MAE, RMSE and SSIM metrics show that our method is the clear winner, the performance improvements with respect to the second-best model seem rather small. To better assess the performance differences among the generative models, we turn our attention to the subjective evaluation study based on the annotations (votes) provided by three independent experts. On average, the experts spent between 16 and 29 hours casting the votes for each presented case. The corresponding results are presented in Table~\ref{tab_doctors}.

The study shows that our approach was voted as the winner by all three experts, in all style transfer experiments. Remarkably, even if the experts did not know which method produced which image, CyTran gathered more than $75\%$ of the votes from each expert. The lowest percentage recorded by CyTran is for Expert \#1, who rated our solution as being the most convenient in $150$ of $200$ ($75\%$) cases. Moreover, CyTran outperforms U-GAT-IT, the second-best model according to Expert \#1, by a considerable margin of $68\%$. Expert \#3 was most favorable towards our model, voting for CyTran in $176$ of $200$ ($88\%$) cases. In contrast, AttentionGAN, the second-best model in opinion of Expert \#3, obtained only $13$ of $200$ ($6.5\%$) votes. Aside from counting the votes casted by each expert, we also analyze the inter-rater agreement via the Cohen's $\kappa$ coefficient. The average inter-rater agreement is $\kappa=0.81$, with values ranging from $0.76$ to $0.88$. We consider this level of agreement as substantial. In conclusion, all experts agree that CyTran is significantly better than the other state-of-the-art translation models. 

After casting the votes, we revealed the identity of the methods to the medical experts. Upon analyzing the generated outputs once again, all three experts observed that CyTran is the only method that does not produce visual artifacts and does not lose information. When a vote went towards a different method, it was almost always decided based on the fact that the voted competing method produced an image closer to the target one.

\begin{figure*}[!t]
\begin{center}
\centering
\includegraphics[width=1.0\linewidth]{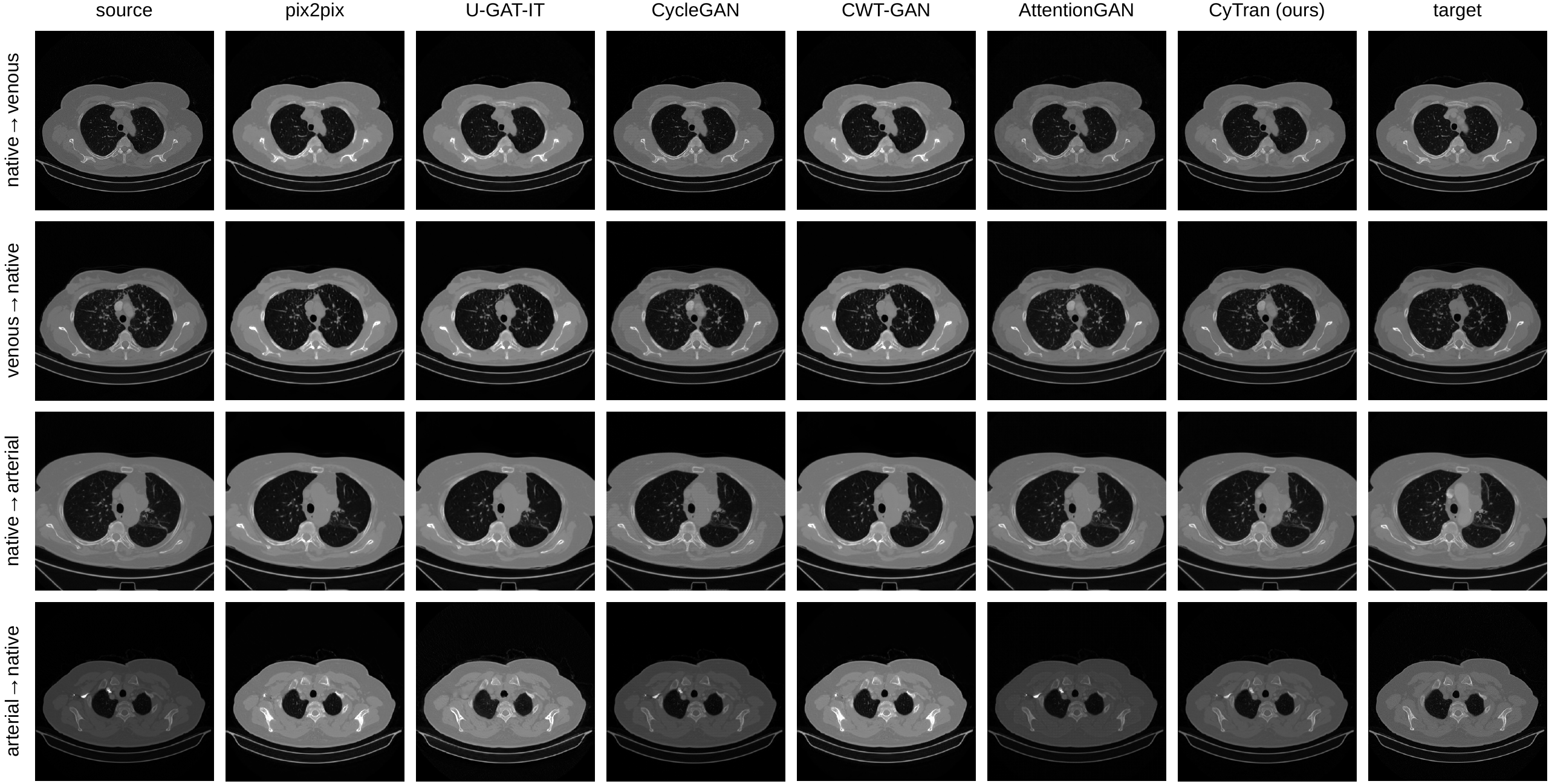}
\caption{Examples of images translated by pix2pix, U-GAT-IT, CycleGAN, CWT-GAN, AttentionGAN and CyTran, respectively. The input (source) and target images are displayed for reference. Defects become more visible when zooming in into this figure.}
\label{fig_examples}
\end{center}
\end{figure*}

\begin{table*}[t]
\renewcommand{\arraystretch}{1.2}
\centering
\noindent
\caption{Volumetric registration results for the venous$\rightarrow$native and arterial$\rightarrow$native contrast pairs. Our model is compared with several state-of-the-art methods and ablated versions of itself on the test set, in terms of MAE and SSIM. The symbol $\uparrow$ means higher values are better, while $\downarrow$ means lower values are better. The best results are highlighted in bold.} 
\begin{tabular}{|l|cc|cc|}
\hline  
\multirow{2}{*}{Method} & \multicolumn{2}{|c|}{{venous$\rightarrow$native}} & \multicolumn{2}{|c|}{{arterial$\rightarrow$native}} \\ 
 & MAE $\downarrow$ & SSIM $\uparrow$ & MAE $\downarrow$ & SSIM $\uparrow$ \\
\hline
\hline
 1-cascade VTN \citep{Zhao-ICCV-2019}    & 0.0098 & 0.6883 & 0.0094 & 0.6894 \\ 
\hline
 3-cascade VTN \citep{Zhao-ICCV-2019}    & 0.0083 & 0.7010 & 0.0081 & 0.7041 \\ 
\hline
 ViT-V-Net \citep{Chen-VITVN-2021}       & 0.0077 & 0.7488 & 0.0071 & 0.7601 \\
 \hline
 ViT-V-Net+pix2pix \citep{Isola-CVPR-2017} & 0.0093 & 0.7210 & 0.0090 & 0.7312 \\ 
\hline
 ViT-V-Net+U-GAT-IT \citep{Kim-ICLR-2019} & 0.0074 & 0.7501 & 0.0070 & 0.7602 \\
 \hline
 ViT-V-Net+CycleGAN \citep{Zhu-ICCV-2017} & 0.0075 & 0.7498 & 0.0069 & 0.7641 \\ 
 \hline
 ViT-V-Net+CWT-GAN \citep{Lai-ICCV-2021} & 0.0073 & 0.7499 & 0.0067 & 0.7644 \\ 
 \hline
 ViT-V-Net+AttentionGAN \citep{Tang-TNNLS-2021} & 0.0073 & 0.7501 & 0.0066 & 0.7667 \\ 
\hline
 ViT-V-Net+CyTran (no MLCC) & 0.0071 & 0.7521 & 0.0066 & 0.7703 \\ 
\hline
 ViT-V-Net+CyTran (ours) & 0.0069 & 0.7545 & 0.0066 & 0.7709 \\ 
\hline
3-cascade ViT-V-Net (ours)              & 0.0064 & 0.7578 & 0.0053 & 0.7772 \\
\hline
3-cascade ViT-V-Net+CyTran (no MLCC)       & 0.0063 & 0.7640 & \textbf{0.0052} & 0.7910 \\ 
\hline
3-cascade ViT-V-Net+CyTran (ours)       & \textbf{0.0061} & \textbf{0.7697} & \textbf{0.0052} & \textbf{0.7933} \\ 
\hline
\end{tabular}
\label{tab_registration}
\end{table*}

\subsubsection{Qualitative analysis}

In Figure~\ref{fig_examples}, we present a randomly sampled case for each of the four contrast pairs. We observe that pix2pix suffers from visual artifacts such as discontinuity in the scapula or erasure of the cortex of the rib, as seen in the native$\rightarrow$arterial translation. Moreover, soft tissues have an increase in noise and there is no additional useful information regarding the vessel contrast. When converting from arterial to native using pix2pix, the structure appears severely altered: the rib is completely dissociated, muscle margins are modified, vessel differentiation is difficult.
Similar to the pix2pix method, U-GAT-IT creates a false warping aura around the bones. The quality of the pulmonary parenchyma is not significantly altered and the appearance of the vessels and soft tissues are harder to differentiate.
CycleGAN increases noise especially around high contrast areas, as seen in the native$\rightarrow$arterial translation. The soft tissues adjacent to bony structure, such as the scapula or the ribs, have a halo of structural deformation. This method also creates a grid-like texture on top of the CT image, which distorts both the structural integrity and increases the number of visual artifacts.
CWT-GAN usually renders a very bright image of the CT scan while losing some essential information, as seen in the aortic arch of the native$\rightarrow$arterial case or the mediastinum in the venous$\rightarrow$native case, respectively. Overall, CWT-GAN is similar or slightly better than U-GAT-IT and inferior to both AttentionGAN or CyTran. 
When comparing our technique with AttentionGAN in the various cases exemplified in Figure~\ref{fig_examples}, we observe that there are a multitude of similarities: high resolution, enriched information, accurate differentiation between different organs or tissues. Despite this fact, CyTran has slightly lower brightness in arterial or venous targets, but higher brightness and contrast in native targets, thus offering a clearer image. The big vessels of the heart can be more accurately described, especially in the native output images. Furthermore, CyTran does not suffer from loss of information regarding the structures, tissue consistency and margins, as seen in the venous$\rightarrow$native translation of the right breast tumor. The true benefit of CyTran lies in the conversion from native CT scans into generated arterial or venous CT scans, because of the fair amount of information acquired with respect to the native image, even if the vessels do not get as bright as in the reference (target) image. In a nutshell, CyTran offers the best experience in generating artificial images, being closest to the target among all methods, both in native and contrast images.

\subsection{Volumetric Registration Experiments}

\subsubsection{Baselines}

As baselines, we consider three state-of-the-art unsupervised medical image registration methods: a 1-cascade Volume Tweening Network (VTN) \cite{Zhao-ICCV-2019}, a 3-cascade VTN \cite{Zhao-ICCV-2019}, and ViT-V-Net \cite{Chen-VITVN-2021}.


\subsubsection{Performance measures}

Since our data set does not contain any labeled segmentation maps, we consider performance measures suitable for evaluating unsupervised registration methods, which quantify the ability to align moving structures, without damaging the integrity of CT scans. Therefore, we report the MAE and SSIM between the warped scan (the alignment result) and the reference scan.

\subsubsection{Data preprocessing}

We apply the same data preprocessing steps as for the style transfer experiments. 

\subsubsection{Hyperparameter tuning}

We train all networks from scratch with the hyperparameters indicated by the authors of cascaded VTNs \cite{Zhao-ICCV-2019} and ViT-V-Net \cite{Chen-VITVN-2021}, except for the mini-batch size, which we reduce to two data samples such that we can train each model on a single GPU.

For our approach, we introduce two hyperparameters: the augmentation rate and the number of cascade steps. The augmentation rate represents the percentage of training data processed with a style transfer method and added to the training set. We consider augmentation rates ranging from $10\%$ to $100\%$, at a step of $10\%$. For the number of cascade steps, we consider all values between $1$ and $4$. We tune these hyperparameters on the validation set, reporting the test results for the optimal configuration found on the validation set.

\begin{figure}[!t]

\captionsetup[subfigure]{labelformat=empty}
\centering
\subfloat[]
{	
    \hspace{-0.23cm}
	\includegraphics[width=0.33\linewidth]{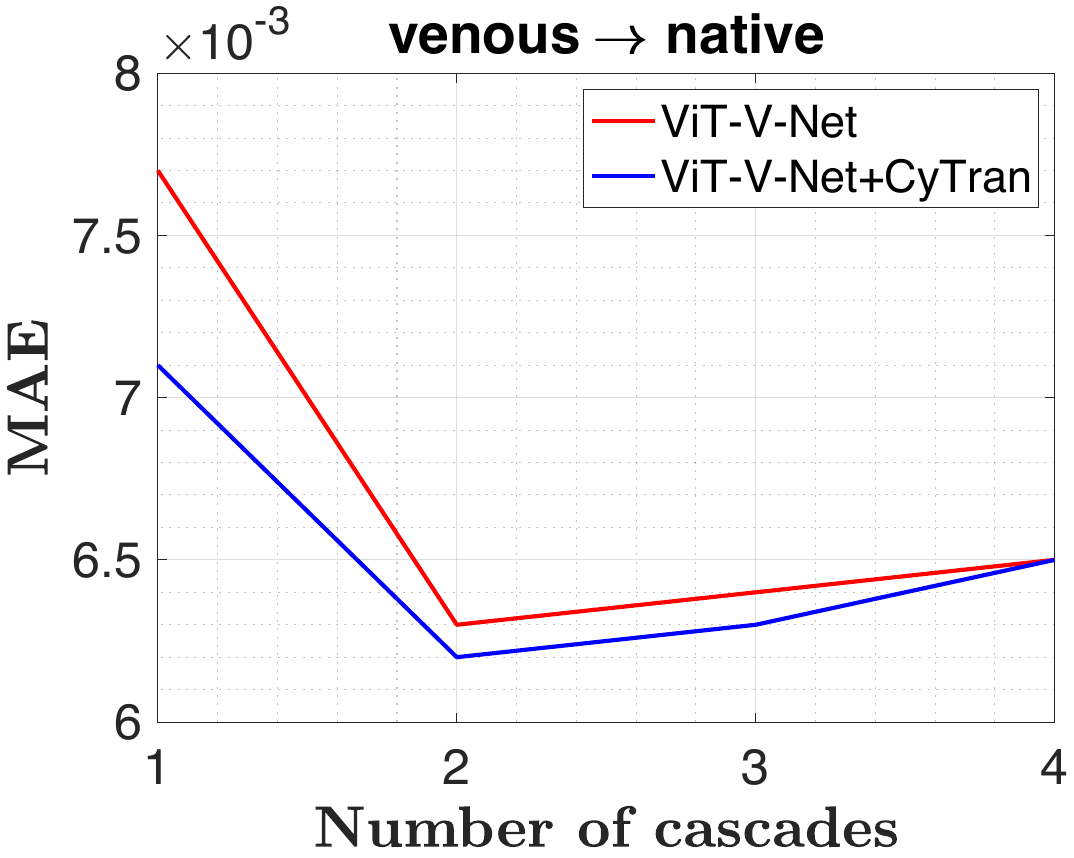}
}
\subfloat[]
{
    \hspace{-0.23cm}
	\includegraphics[width=0.33\linewidth]{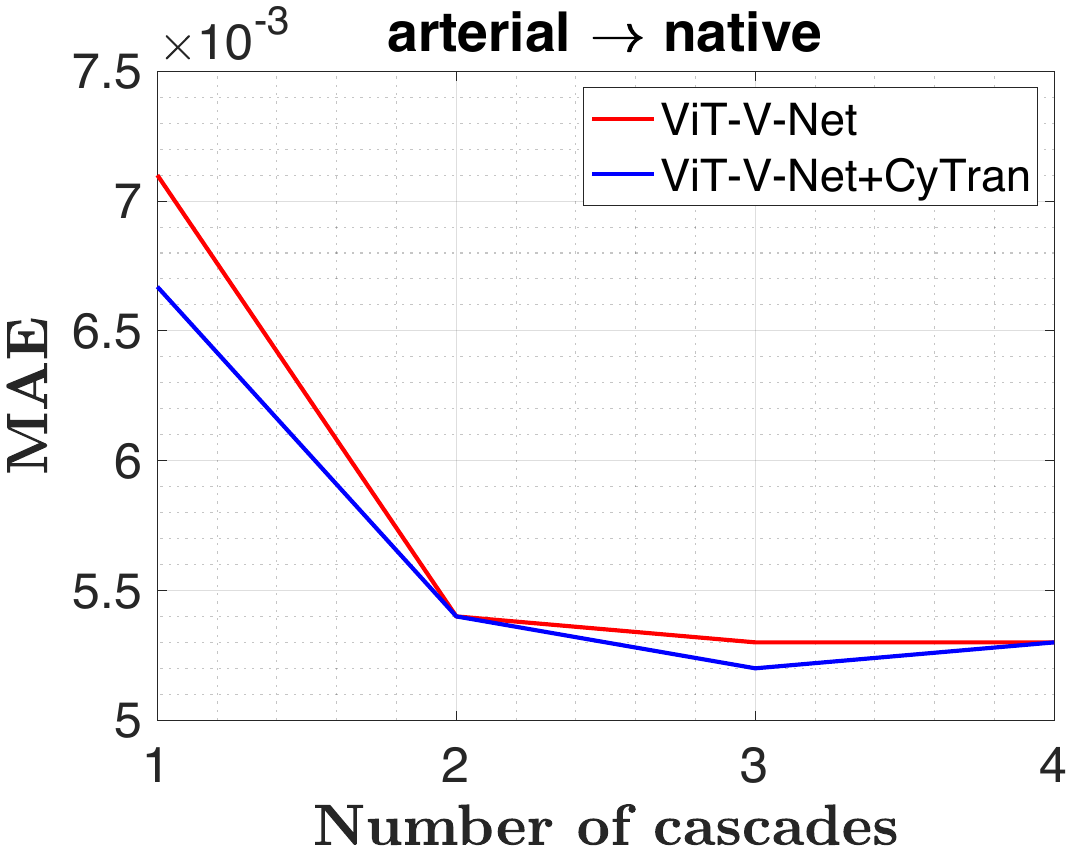}
}
\\
\vspace{-0.5cm}
\subfloat[]
{	
    \hspace{-0.23cm}
	\includegraphics[width=0.33\linewidth]{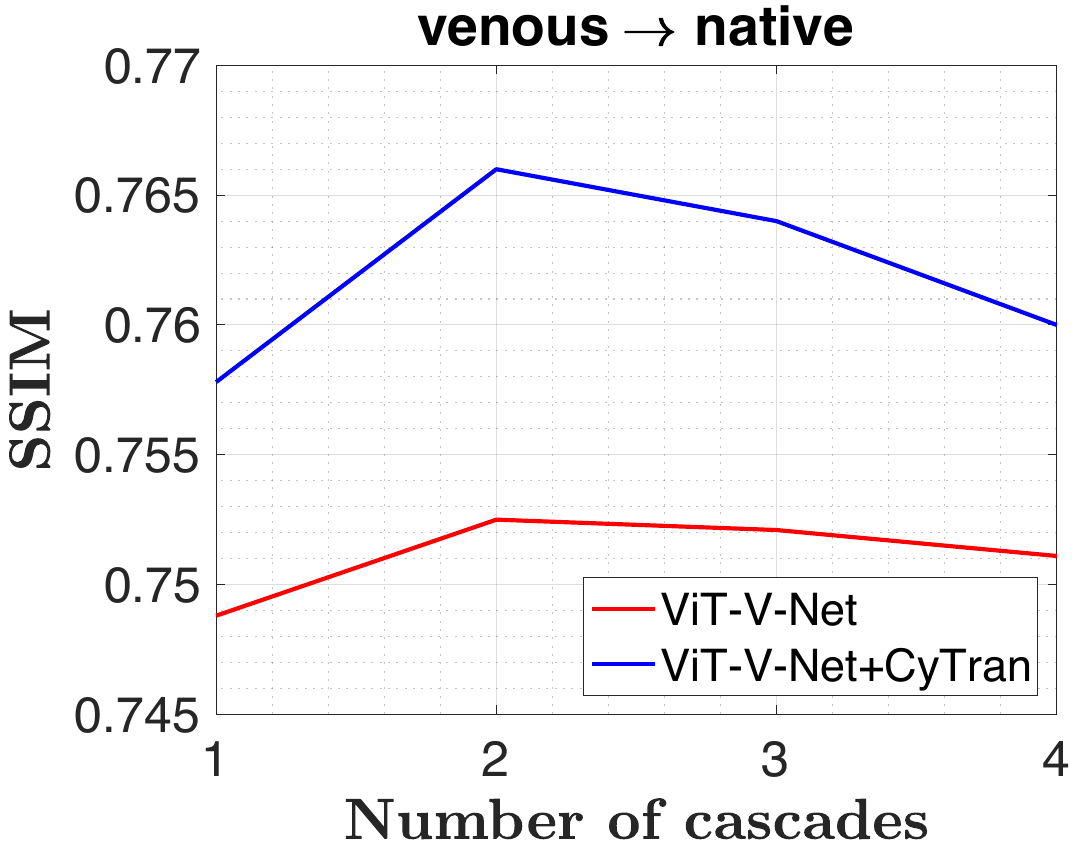}
}
\subfloat[]
{
    \hspace{-0.23cm}
	\includegraphics[width=0.33\linewidth]{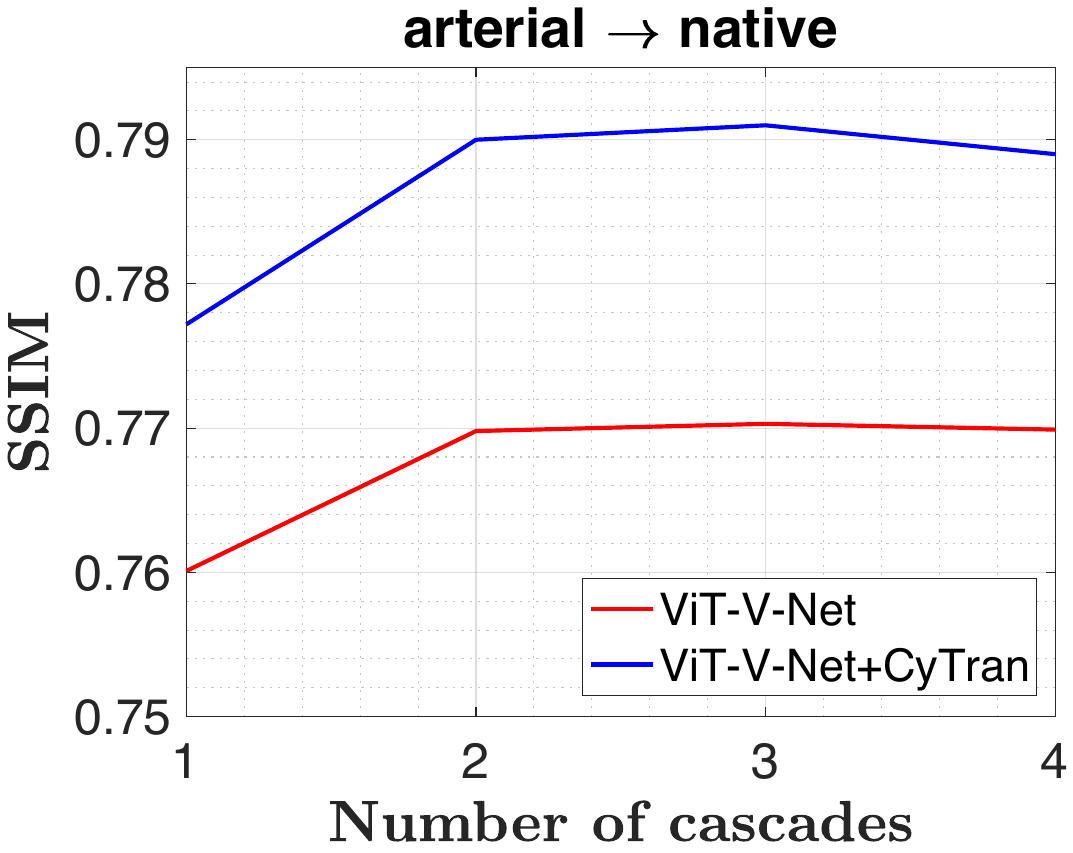}
}
\caption{Results for cascaded ViT-V-Net \citep{Chen-VITVN-2021} and ViT-V-Net+CyTran models with various cascade steps. Models without recursive cascade are equivalent to models with one cascade step.\label{fig_registration_cascade}}
\end{figure}

\begin{figure}[!t]
\begin{center}
\centering
\includegraphics[width=1.0\linewidth]{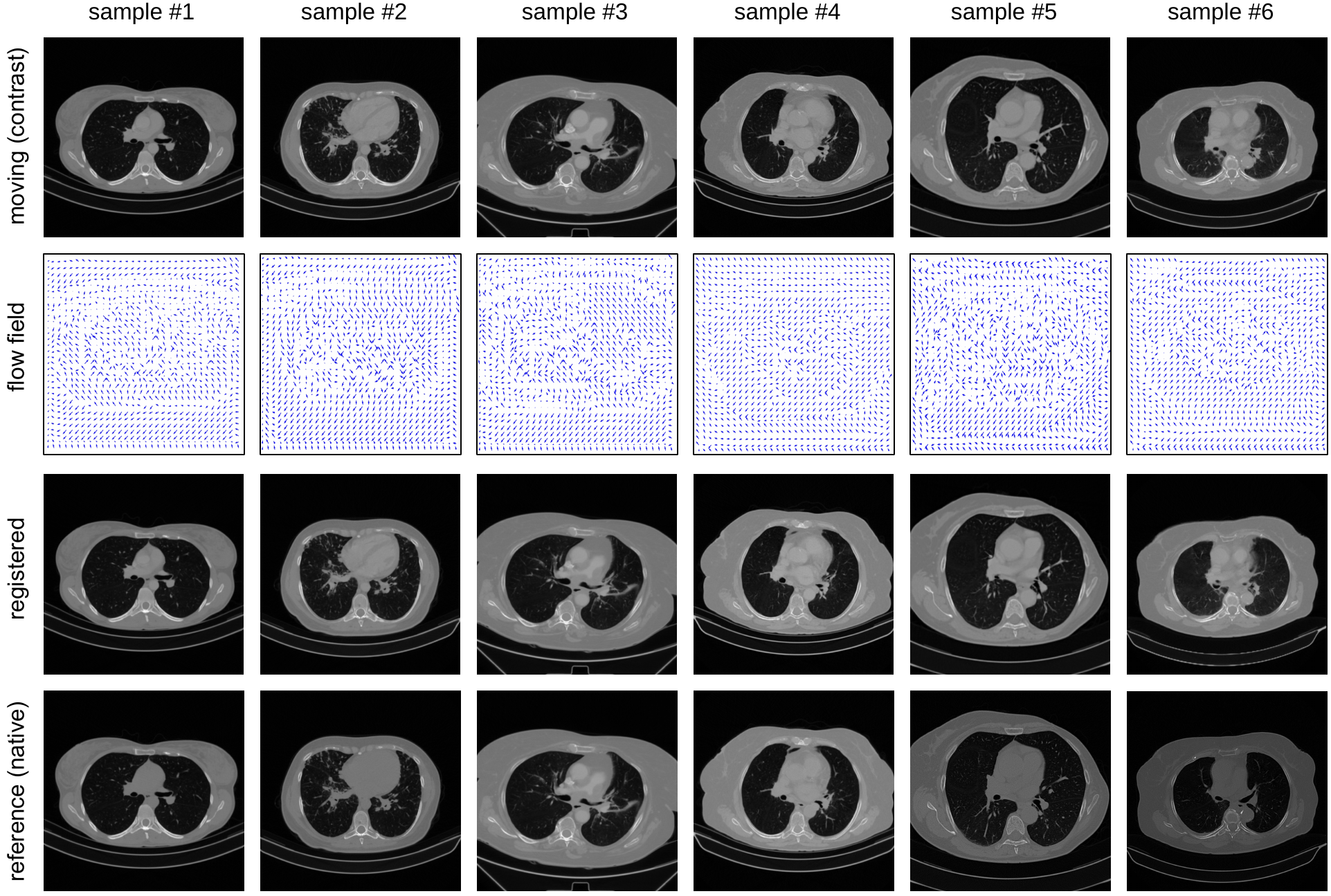}
\caption{Examples of images registered by our best model, namely the 3-cascade ViT-V-Net+CyTran. The first three examples illustrate the registration of arterial to native slices, while the last three examples show the registration of venous to native slices. For each moving image (arterial or venous), we present the displacement map (or flow field) and the registered images (obtained by applying the flow field to the moving images). In all cases, the reference image is native (without contrast).}
\label{fig_examples_align}
\end{center}
\end{figure}

\subsubsection{Quantitative results}

In Table~\ref{tab_registration}, we present the comparative results between our 3-cascade ViT-V-Net framework trained on images translated by CyTran and the three state-of-the-art methods, two based on cascaded VTNs \cite{Zhao-ICCV-2019} and one based on ViT-V-Net \cite{Chen-VITVN-2021}. In addition, we show ablation results obtained by removing the cascade, CyTran, or the multi-level cycle-consistency (MLCC) from our model. We also report results for the ViT-V-Net model trained on images translated by our competitors: pix2pix, U-GAT-IT, CycleGAN, CWT-GAN and AttentionGAN.

First, we observe that training ViT-V-Net on images translated by pix2pix \cite{Isola-CVPR-2017} damages performance, leading to worse results for both venous$\rightarrow$native and arterial$\rightarrow$native pairs, in comparison with the vanilla ViT-V-Net \cite{Chen-VITVN-2021}. In contrast, U-GAT-IT \cite{Kim-ICLR-2019}, CycleGAN \cite{Zhu-ICCV-2017}, CWT-GAN \cite{Lai-ICCV-2021}, AttentionGAN \cite{Tang-TNNLS-2021} and CyTran bring performance gains, indicating that our idea of transferring the style of source CT scans to target CT scans before registration is useful. Among these three models, CyTran gives the highest performance gains, once again showing its superiority over pix2pix, U-GAT-IT, CycleGAN, CWT-GAN and AttentionGAN.

Interestingly, our idea of introducing ViT-V-Net into a recursive cascade is also useful. This observation is confirmed by the fact that the 3-cascade ViT-V-Net outperforms ViT-V-Net, as well as the fact that the 3-cascade ViT-V-Net+CyTran outperforms ViT-V-Net+CyTran. To further confirm our observation, we present results for ViT-V-Net and ViT-V-Net+CyTran with various numbers of cascade steps in Figure~\ref{fig_registration_cascade}. We observe that having more than one cascade brings considerable improvements over the baselines. For both approaches, the highest gains are obtained with $3$ cascades for the arterial$\rightarrow$native pair, and $2$ cascades for the venous$\rightarrow$native pair.

In the style transfer experiments, we did not observe major differences between the multi-level cycle-consistency (MLCC) loss and the standard cycle-consistency loss for the native target domain. However, even minor differences in style transfer could lead to better alignment. To put this statement to the test, we included an ablated version of ViT-V-Net+CyTran, where the MLCC loss is replaced with the standard cycle-consistency loss. The reported results (shown in Table~\ref{tab_registration}) indicate that the MLCC loss leads to superior alignment performance, confirming that our MLCC loss is a useful contribution, even for the native target domain.

In summary, the empirical results show that our recursive cascaded ViT-V-Net based on style transfer with CyTran is the best approach for non-contrast to contrast CT scan registration, surpassing all baselines and ablated models.

\subsubsection{Qualitative analysis}

In addition to our quantitative results, we illustrate six randomly chosen registration results of our best registration model in Figure \ref{fig_examples_align}. In the first three columns, the phase of the moving image is arterial, while in the last three columns, the phase is venous. First, we observe that the moving images (arterial or venous) are not far from being well aligned to the reference images. However, the contrast substance introduces significant differences between pixel intensities (HU levels) of certain organs, e.g.~the heart ventricles in the second sample or the pulmonary arterial tree vessels in the sixth sample. This observation indicates that a standard registration model (without a style transfer model to help reduce the HU level differences between contrast and non-contrast scans) needs to learn to match pixel intensity differences, while performing the alignment. Moreover, the pixel intensity differences also interfere with the loss function, creating large loss values even if the slices are well aligned. We alleviate these issues by introducing CyTran into the registration pipeline, significantly easing the task for the ViT-V-Net registration model. Indeed, the registered (output) slices depicted in Figure \ref{fig_examples_align} confirm that the proposed registration pipeline works well.

\section{Conclusions}

In this paper, we introduced cycle-consistent convolutional transformers in medical imaging. We employed our approach to transfer the style between contrast and non-contrast CT scans, showing that it outperforms state-of-the-art methods such as pix2pix, U-GAT-IT, CycleGAN, CWT-GAN and AttentionGAN. Our qualitative and subjective human evaluations revealed that CyTran is the only approach that does not introduce visual artifacts during the translation process. We believe this is a key advantage in our application domain, where medical images need to precisely represent the scanned body parts. Moreover, we showed that CyTran brings significant improvements for a state-of-the-art medical image registration method. Another important contribution of our work is Coltea-Lung-CT-100W, a new data set of triphasic CT scans comprising a total of 37,290 images. In future work, we aim to apply our registration results to improve multi-image super-resolution and lesion segmentation.

\section*{Acknowledgements}

The research leading to these results has received funding from the NO Grants 2014-2021, under project ELO-Hyp contract no.~24/2020. This article has also benefited from the support of the Romanian Young Academy, which is funded by Stiftung Mercator and the Alexander von Humboldt Foundation for the period 2020-2022.





\bibliographystyle{elsarticle-num}
\bibliography{ref}







\end{document}